\documentclass[ letterpaper,twocolumn,10pt]{article}
\usepackage{usenix2019}
\AtBeginDocument{%
  \providecommand\BibTeX{{%
    \normalfont B\kern-0.5em{\scshape i\kern-0.25em b}\kern-0.8em\TeX}}}

\usepackage{algorithmicx}
\usepackage{algpseudocode}
\usepackage{amsmath}
 \usepackage{etoolbox}
\makeatletter
\patchcmd{\maketitle}{\@copyrightspace}{}{}{}
  \usepackage{booktabs} % For formal tables
\usepackage{float}

\usepackage{amsmath}
\usepackage{diagbox}
\usepackage{color}
\usepackage[utf8x]{inputenc}
\usepackage{indentfirst}
\usepackage{hyperref}
\usepackage{subfigure}
 \usepackage{graphicx}
 \usepackage{graphicx}
\usepackage{color}
\usepackage{xcolor}
\usepackage[numbers,sort&compress]{natbib}

\usepackage{url}

\usepackage{graphics}
\usepackage{framed} 
\usepackage{listings}
\usepackage{pifont}
\definecolor{codegreen}{rgb}{0,0.6,0}
\definecolor{codegray}{rgb}{0.5,0.5,0.5}
\definecolor{codepurple}{rgb}{0.58,0,0.82}
\definecolor{backcolour}{rgb}{0.95,0.95,0.92}
\renewcommand{\paragraph}[1]{\vspace{0.12in}\noindent{\bf{#1}.}}

\newcolumntype{P}[1]{>{\centering\arraybackslash}p{#1}}
\newcommand{\ignore}[1]{}
\usepackage[T1]{fontenc}
 
\usepackage{enumerate}
\usepackage{authblk}

\newenvironment{packed_item}{
\begin{itemize}
  \setlength{\itemsep}{1pt}
  \setlength{\parskip}{0pt}
  \setlength{\parsep}{0pt}
}{\end{itemize}}

\begin{document}

\vspace{-10mm}

\title{On the (In)security of Bluetooth Low Energy  One-Way \\ Secure Connections Only Mode }

%\title{One-Way Secure Only is Insecure: \\On Enforcing Secure Pairing in Bluetooth Low Energy with Mobile Devices}

\author[$\Delta$,$\gamma$]{Yue Zhang }
\author[$\Delta$]{Jian Weng}
\author[$\gamma$]{Rajib Dey}
\author[*]{Yier Jin}
\author[**]{Zhiqiang Lin}
\author[$\gamma$]{Xinwen Fu}

\affil[$\gamma$]{Department of Computer Science,~University of Central Florida}
\affil[$\Delta$]{College of Information Science and Technology,~ Jinan University}
\affil[*]{Department of Electrical \& Computer Engineering,~ University of Florida}
\affil[**]{Computer Science and Engineering, The Ohio State University}

\ignore{
\author{Yue Zhang}
\affiliation{%
  \institution{College of Information Science and Technology,Jinan University}
  \streetaddress{}
 \city{GuangZhou}
  \state{GuangDong}
  \postcode{}
}
\author{Jian Weng}
\affiliation{%
  \institution{College of Information Science and Technology,Jinan University}
  \streetaddress{}
  \city{GuangZhou}
  \state{GuangDong}
  \postcode{}
}
\author{Rajib Dey}
\affiliation{%
  \institution{Department of EECS, University of Central Florida}
  \streetaddress{}
  \city{Orlando}
  \state{Florida}
  \postcode{}
}
\author{Yier Jin}
\affiliation{%
  \institution{University of Florida}
  \streetaddress{}
  \city{Gainesville}
  \state{Florida}
  \postcode{}
}
\author{Zhiqiang Lin}
\affiliation{%
  \institution{Computer Science and Engineering, The Ohio State University}
  \streetaddress{}
  \city{Columbus}
  \state{Ohio}
  \postcode{}
}
\author{Xinwen Fu}
\affiliation{%
  \institution{Department of EECS, University of Central Florida}
  \streetaddress{}
  \city{Orlando}
  \state{Florida}
  \postcode{}
}
}

%\keywords{Bluetooth Low Energy, Secure Connection Only Mode, Android Security, Downgrade Attack, Enforce Secure Pairing}

% \settopmatter{printacmref=false} % Removes citation information below abstract
%\renewcommand\footnotetextcopyrightpermission[1]{} % removes footnote with conference information in first column
\pagestyle{plain} % removes running headers
\maketitle

\begin{abstract} \noindent
To defeat security threats such as man-in-the-middle (MITM) attacks, Bluetooth Low Energy (BLE) 4.2 and 5.x introduce the Secure Connections Only mode, under which a BLE device accepts {\em only} secure paring protocols including Passkey Entry and Numeric Comparison from an initiator, e.g., an Android mobile. 
However, the BLE specification does not explicitly require the Secure Connection Only mode of the initiator. Taking the Android's BLE programming framework for example, we found that it cannot enforce secure pairing, invalidating the security protection provided by the Secure Connection Only mode. The same problem applies to Apple iOS too.

Specifically, we examine the life cycle of a BLE pairing process in Android and identify four severe design flaws. 
These design flaws can be exploited by attackers to perform downgrading attacks, forcing the BLE pairing protocols to run in the insecure mode without the users' awareness.
To validate our findings, we selected and tested 18 popular BLE commercial products and our experimental results proved that downgrading attacks and MITM attacks were all possible to these products.
All 3501 BLE apps from Androzoo are also subject to these attacks.
For defense, we have designed and implemented a prototype of the Secure Connection Only mode on Android 8 through the Android Open Source Project (AOSP).
We have reported the identified BLE pairing vulnerabilities to Bluetooth Special Interest Group (SIG), Google, Apple, Texas Instruments (TI) and all of them are actively addressing this issue. Google rated the reported security flaw a High Severity.
\end{abstract}

\section{Introduction}

Bluetooth Low Energy (BLE) is a widely adopted wireless communication technology and is broadly adopted in IoT such as medical applications including blood pressure monitoring and X-ray imaging as well as wearable technologies. \cite{BSIG::Market::2019}.
BLE has two salient features: low energy consumption to increase the lifetime of battery-powered BLE devices and a development framework - GATT (Generic Attribute Profile) to allow mobile, tablet and PC applications for arbitrary data transmission to peer BLE devices. \looseness=-1

BLE relies on a pairing protocol to ensure the communication security. Two pairing devices authenticate each other and negotiate a secret key to encrypt the communication channel. 
To achieve this goal, latest versions of the specification including Bluetooth 4.2 \cite{4.2} and 5.x \cite{5.0, 5.1} offer four pairing protocols: 
Just Works, Passkey Entry, Numeric Comparison and Out Of Band ({\em OOB}). Just Works is not secure and OOB is rarely used due to the extra hardware cost.  
Therefore, we denote Passkey Entry and Numeric Comparison as practical secure pairing protocols and will focus on the security analysis of these two pairing protocols.

The latest BLE 4.2 \cite{4.2} and 5.x \cite{5.0} add the new Secure Connections Only mode for BLE enabled devices to address vulnerabilities found in previous Bluetooth pairing protocols \cite{xu2019badbluetooth,lonzetta2018security,melamed2018active,gupta2016inside,jasek2016gattacking,uher2016denial,kwon2016bluetooth,o2015security,zegeye2015exploiting,NaveedZDWG14,cyr2014security,ryan2013bluetooth,rosa2013bypassing,sandhya2012analysis,haataja2010two,haataja2008practical}. 
Man-in-the-middle (MITM) attacks in \cite{haataja2010two,haataja2008practical} work against Bluetooth Secure Simple Pairing (SSP) of Bluetooth Classic 2.1 and 3.0, in which two Bluetooth devices under SSP use only I/O capabilities (such as display and keyboard) to determine the pairing protocol. 
An attacker can falsely declare their I/O capabilities and conduct an MITM attack \cite{haataja2011ten,almomani2011secure,bai2016staying}. 
%The BLE GATT profile often involves a mobile device such as an Android mobile and a (peer) BLE device. The mobile device and its peer BLE device exchange the I/O capabilities and other pairing features to determine the pairing protocol. 
With BLE 4.2 and 5.x, if the Secure Connections Only mode is enabled in a BLE device, the BLE device is forced to authenticate the user/mobile device with secure pairing protocols. It is expected that the Secure Connections Only mode will enforce secure pairing to defeat the MITM attack. 

However, BLE does not explicitly specify such a Secure Connections Only mode for a connection initiator such as a mobile device (more details are available in Section \ref{subsec::SecureConnectionsOnly}). 
Without the Secure Connections Only mode being enforced at the initiator side, the initiator, e.g., a mobile, is not required to authenticate the BLE device.
%In this paper, we first investigate the security impacts and potential vulnerabilities because there is no Secure Connections Only mode for mobile devices. 
We find such vulnerabilities exist in both Android and iOS systems and believe that this problems is a protocol level issue. That is, all BLE systems following the BLE specification will have this problem. In this paper, however, we will focus on the Android system with brief discussion on the iOS system due to the page limit.

After examining the initiation, status management, error handling and bond management throughout the life cycle of a pairing process in the Android programming framework, we identify four design flaws which, if not properly addressed, cannot enforce the Secure Connections Only mode.
\begin{packed_item}
\item An Android app cannot specify any pairing protocol even if it knows its peer BLE device's capable pairing protocols;
\item An Android app cannot cancel an insecure pairing process until the pairing process is completed;
\item A fake device may poke an Android device and intentionally create pairing errors. Android mishandles those pairing errors without notifying the app and the user;
\item Even if an app knows insecure pairing is used, Android does not provide a mechanism to remove the previously generated key or start a new secure pairing process with the specific peer BLE device.
\end{packed_item}

The four Android BLE design flaws cause serious security issues. 
For example, if an Android mobile was paired with a peer BLE device through secure pairing protocols, then a fake/spoofing device can downgrade the pairing protocol, pair with an Android mobile using insecure pairing protocol (i.e., Just Works) or even communicate in plaintext and inject data into the Android device and the corresponding BLE app. Note that the Identity Resolving Key (IRK) is designed to prevent a mobile from leaking its MAC address and thus being tracked. With these security flaws, a fake device can pair with the victim mobile using Just Works and obtain the mobile's IRK and MAC address to track the mobile device. Even worse, the fact that Android cannot enforce secure pairing causes damages beyond the mobile itself. A BLE device may implement a whitelist that allows only previously paired mobile devices to connect or to access services. An attacker can now collect the mobile device's MAC address and IRK to bypass the whitelist-based filtering.

To solve these security issues, we believe that the Android system shall be able to enforce the Secure Connections Only mode such that a BLE app can specify and enforce a secure pairing protocol. 
%\red{ Android should also notify the user if any pairing error occurs and grant the user the permission to connect/disconnect suspicious BLE devices. }
If secure pairing is enforced on Android, a mobile device user has to see pairing BLE devices and make the decision physically to disable/proceed with the pairing process. If the pairing device is a fake device, the fake device
will be identified by the user and any following attacks shall fail. The secure connections only mode at the app side will not create any conflict with peer devices. Practically, the app
and its peer device know each other’s capabilities and shall be able to enforce the secure connections only mode if secure pairing is desired and their I/O capabilities support it. We advocate the option of the secure connections only mode for the app so that the app can defeat various attacks. If the option is not used, BLE shall run compatibly. 

Our major contributions are summarized as follows.
\begin{packed_item}
\item For the first time, we find that BLE does not require the Secure Connections Only mode for a pairing initiator such as a mobile phone. 
The lack of mutual authentication with Secure Connections Only mode at both the mobile and its peer device in the BLE specification causes security vulnerabilities. 
%If secure pairing is needed to protect sensitive data and communication, manufacturers shall equip their devices with such I/O capacities. 
%\vspace{-2mm}
\item Based on our finding on BLE specification security flaws, we tested and proved that both Android and iOS do not provide a programming framework for BLE apps to enforce security pairing. 
%implement the Secure Connections Only mode, enforcing secure pairing. 
Specifically, four design flaws were identified in Android leading to security vulnerabilities on Android mobiles and peer BLE devices. 
%\vspace{-2mm}
\item  Thorough experimentation was performed on BLE apps running on the latest versions of Android and 18 commercial BLE devices. Not surprisingly, all 3501 BLE apps from Androzoo \cite{allix2016androzoo} are subject to downgrading and MITM attacks. In our experiments, the line-of-sight attacking distance can reach 76 meters.
%\vspace{-2mm}
\item  Security defenses and solutions are proposed and prototyped to enhance the Secure Connections Only mode for Android by enforcing secure pairing protocols through the Android Open Source Project (AOSP) \cite{AOSP}. Our case study on BLE keyboards further prove that the Numerical Comparison protocol is more secure than the Passkey Entry protocol even if both the mobile and the peer device enforce secure pairing. Therefore, for mission critical BLE apps and devices, we suggest that Numerical Comparison will be used on the mobile and on BLE devices to provide higher security assurance. 
%The mechanisms we introduce to defeat attacks may require user interaction, but take effect only when there are attacks and will not affect the usability much.
\end{packed_item}

\smallskip\noindent{\bf Responsible Disclosures:} We have reported our findings to Bluetooth Special Interest Group (SIG), Google Android Security Team, Apple, and Texas Instruments (TI) Product Security Incident Response Team (PSIRT). The Bluetooth SIG acknowledged our findings and is currently working with Google to address the issues. The Google Android Security Team also acknowledged the four design flaws  in Section \ref{sec:pitfalls} and rated the identified Android vulnerabilities as {\bf High} severity \cite{severity}. They are actively working with us to patch Android. TI's PSIRT has released a patched SDK to ``Update authentication parameters when transitioning between authenticated/non-authenticated pairing'' based on the reported vulnerabilities of TI's BLE stack \cite{Ti-new-SDK,Tifix}. 

\smallskip\noindent{\bf Roadmap:} The rest of the paper is organized as follows. We introduce BLE in Section \ref{sec:ble} and present the BLE pairing process flaws in Section \ref{sec:pitfalls}. Section \ref{sec:attacks} presents the downgrading attacks exploiting the design flaws. Case studies are presented in Section \ref{sec:casestudy}. 
Section \ref{sec:countermeasure} discusses various countermeasures including the enforced secure paring in Android.
Section \ref{sec:eva} evaluates the attacks and countermeasures. 
Section \ref{sec:discuss} discusses securing pairing on iOS and other potential solutions to BLE security.
Related work is presented in Section \ref{sec:relatedwork} and we conclude the paper in Section \ref{sec:conclusion}. 

%and CC26XX development boards from Texas Instruments (TI) 

\section{Background}% - Bluetooth Low Energy}
\label{sec:ble}

In this section, we will first present an overview of BLE, and then introduce the connection setup process, the pairing process, and the Secure Connections Only mode. We will also introduce the {\em Attribute Protocol (ATT)}.

\subsection{BLE Overview}
\label{sec:bleoverview}

\begin{figure}
\flushright
%\vspace{-2mm}
\includegraphics[width=0.9\columnwidth]{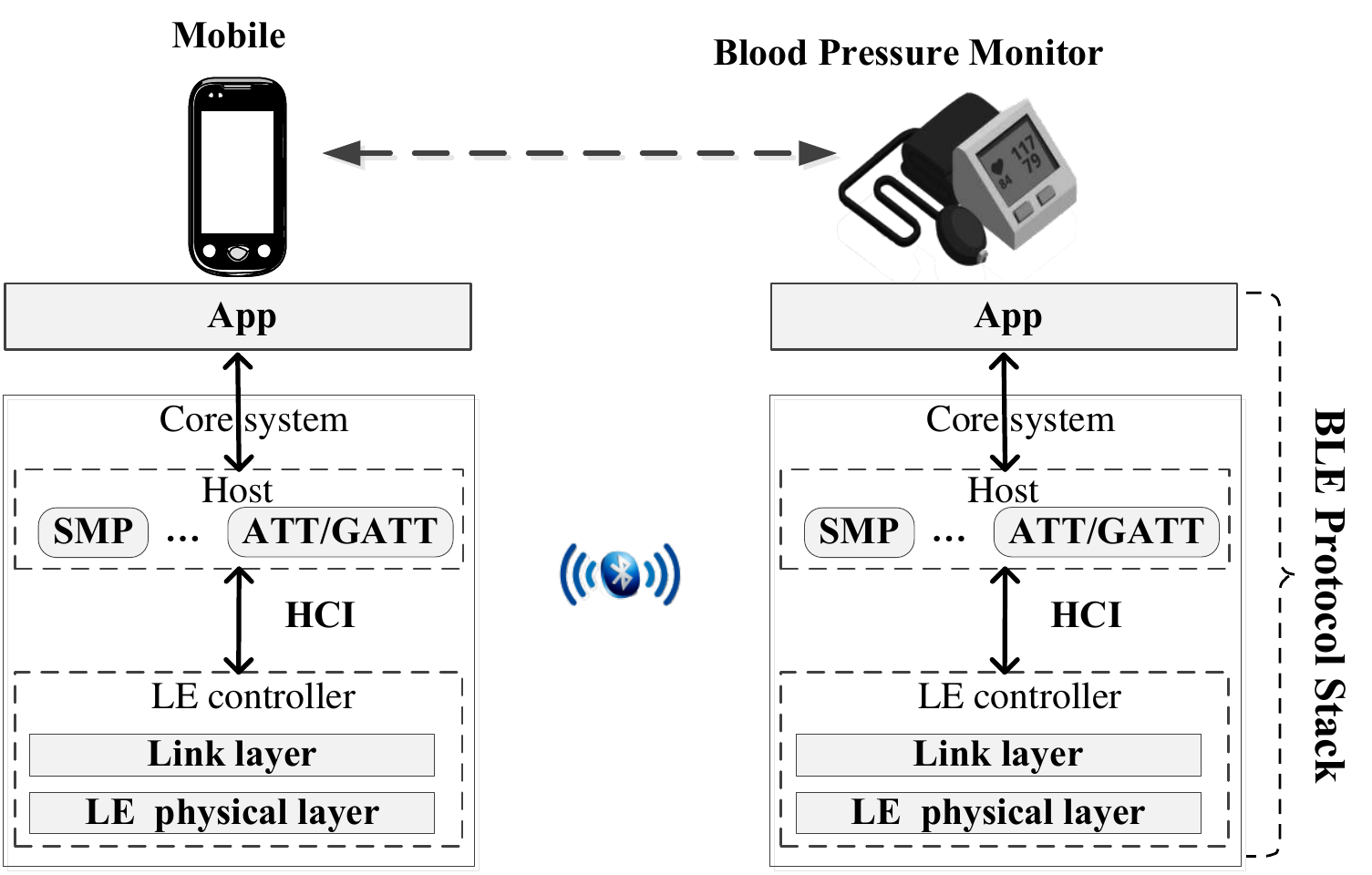} 
\caption{BLE protocol stack}\label{fig:overview}
\end{figure}

BLE is a short-range communications technology. 
Figure \ref{fig:overview} shows its protocol stack where a blood pressure monitor is used as exemplary BLE device. 
In this example, the application on the blood pressure monitor measures blood pressure. The blood pressure monitor application and the mobile app use the {\em BLE core system} for communication. 
BLE's core system consists of two building blocks, LE controller and host.
The LE controller uses the link layer and physical layer to create a connection for sending/receiving data. BLE's physical layer uses frequency hopping for communication, where data is exchanged over a sequence of hopping frequencies. The frequency hopping sequence is negotiated between two devices. 
The host implements multiple protocols including the Security Manager Protocol (SMP) and ATT for secure communication over the connection. 
ATT is used to format the transmitted data.
SMP uses pairing protocols to negotiate cryptographic keys for data encryption, integrity and other purposes. 
The Host Controller Interface (HCI) moves the data, e.g., blood pressure measurements or SMP control commands, from the host to the LE controller through a physical interface, a function call or other venues depending on specific implementations.
 
%\vspace{-2mm}
\subsection{Connection Setup}
\label{sec:con}
%\vspace{-1mm}

\begin{figure} 
%\vspace{-2mm}
\includegraphics[width=\columnwidth]{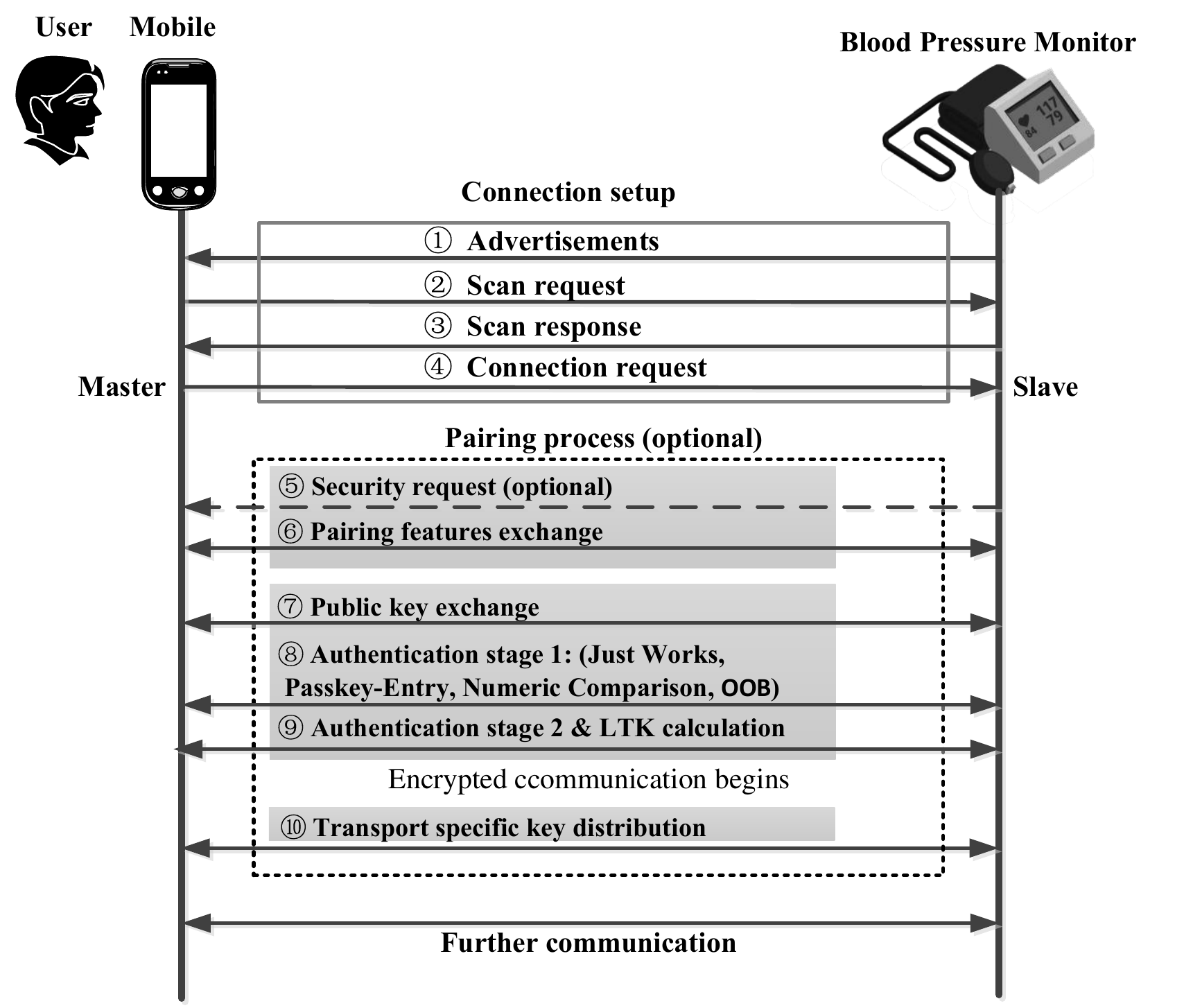}
%\vspace{-2mm}
\caption{Bluetooth Low Energy (BLE) communication}\label{fig:states}
%\vspace{-2mm}
\end{figure}

Steps 1 to 4 in Figure \ref{fig:states} illustrate a typical BLE connection setup process.
Exact information exchanged at each step varies based on different applications.
In Step 1, the blood pressure monitor broadcasts advertising packets indicating its availability. When a mobile app is launched, the app utilizes the Host and receives the advertisements. 
In Step 2, the mobile app sends a scan request to the monitor.
In Step 3, the blood pressure monitor responds with a scan response packet.
The mobile app uses advertising and scanning to collect information about the blood pressure monitor such as the monitor's name, the MAC address and primary services.
In Step 4, the mobile app can now decide if the device is the one of interest and send the connection request to build the connection. The frequency hopping increment is included in the connection request which determines the frequency hopping sequence that the mobile and the blood pressure monitor will follow in the communication. Here the mobile is called the master/initiator for its role in initiating the connection. The peer BLE device, the blood pressure monitor in this case, is called the slave/responder.

%\vspace{-2mm}
\subsection{Pairing Process}
\label{sec:pair}
%\vspace{-1mm}

After two BLE devices build a connection, if no device explicitly requests pairing, the communication continues in plaintext. The two devices need to explicitly start the pairing process to negotiate keys and encrypt the communication. Steps 5 to 9 in Figure \ref{fig:states} illustrate a typical pairing process. A mobile app can initiate the pairing process through SMP (see Figure \ref{fig:overview}). The end-user can also use the system setting app to start the pairing process.
 
\vspace{-2mm}
\subsubsection{Phase 1 - Pairing feature exchange}
\label{sec:pairfeature}
\vspace{-1mm}

Step 5 - Security request (optional). 
As a slave device, the blood pressure monitor can send a security request and ask the mobile (master) to initiate the pairing process. 

Step 6 - Pairing feature exchange. The mobile app sends out a pairing request and the blood pressure monitor returns a pairing response. The two devices then announce their Input/Output (I/O) capabilities such as keyboard and display, authentication requirements and the BLE version so that a suitable common pairing protocol can be negotiated. 
(i) The authentication requirements can be {\em bonding} and {\em MITM protection}. Bonding means that the keys generated during the pairing process will be saved for later use to reduce delay caused by the future pairing process. If a device wants to defend against the MITM attack, the MITM flag \cite{4.2-p3} must be specified so that the Passkey Entry protocol or the Numeric Comparison protocol will be adopted.
The exchanged I/O capabilities will help select communication protocol since different pairing protocols require different I/O capabilities. For example, Numeric Comparison requires a display on both devices. 
If authentication requirements are requested but I/O features cannot support the specified secure pairing, according to the BLE specification, the communication shall be terminated and the user will be notified.
(ii) If the two devices explicitly set the MITM flag as false, Just Works is selected and Just Works cannot defend against the MITM attack. 
(iii) The BLE version is indicated in the {\em Secure Connections (SC)} bit. If the mobile and peer device set the SC bit, BLE 4.2 and above will be adopted. Otherwise, the BLE legacy pairing method is used. We focus on only BLE 4.2 and above in this paper.

\vspace{-2mm}
\subsubsection{Phase 2 - Key exchange and authentication}
\label{subsubsec:pairprotocols}
\vspace{-1mm}

Step 7 - Public key exchange. The Elliptic-curve Diffie–Hellman (ECDH) key exchange protocol is performed so that master and slave devices obtain each other's public key and generate a symmetric key, DHKey.

Step 8 - Authentication stage 1 (a critical step to BLE security). 
In this step, authentication related information is exchanged between the two devices. A pin is entered if Passkey Entry is used. A 6-digit number is displayed at the two devices if Numeric Comparison is used. A verification procedure will also run to make sure the public keys exchanged in Step 7 are from the intended devices.

Step 9 - Authentication stage 2 and LTK calculation. 
%This step is the same for all pairing protocols. 
The previously exchanged authentication information including DHKey is used to generate the MacKey and Long Term Key (LTK) at the two pairing devices. MacKey is used in a process to ensure both devices generate the same LTK. If bonding is required in Step 6, LTK is saved for future session key generation and link encryption.
BLE defines two types of keys, unauthenticated-and-no-MITM-protection keys for Just Works and authenticated-and-MITM-protection keys for Passkey Entry, Numeric Comparison and OOB. 

To help readers better understand these communication protocols, we briefly introduce them below with more details in \cite{4.2}.

\smallskip{\bf Passkey Entry}: During the pairing process, one device such as a mobile needs to display a 6-digit pin, and the user inputs the pin on the other device using a keypad/keyboard. The authentication stage 1 in Step 8 will fail if the attacker does not know the pin.

\smallskip{\bf Numeric Comparison} (BLE 4.2 and beyond): Numeric Comparison is applicable when both devices have displays and confirmation buttons. 
After the ECDH key exchange, the two BLE devices exchange a pair of nonces in Step 8. A function is then used to convert the exchanged public keys and nonces into a six-digit number. Each device displays the number \cite{Numeric-Comparison}. The user confirms that the two displayed numbers are the same by pressing the ``Yes'' button on each device's display to proceed the pairing process. The fact that the two displayed numbers are the same ensures that the exchanged two pubic keys are from the two intended pairing devices, other than from an an attacker. 

\smallskip{\bf Out of Band (OOB)}: In {\em OOB}, a secret is shared with an out-of-band venue such as near-field communication (NFC) and the LTK is derived from this secret. If the OOB venue is secure, the MITM attack can be defeated. 

\smallskip{\bf Just Works}: It is designed for devices without I/O capabilities \cite{4.2} and is subject to MITM attacks. Just Works has almost the same pairing process as Numeric Comparison except that the generated number is not displayed and the user has cannot ensure the exchanged pubic keys are the same.\looseness=-1

\vspace{-2mm}
\subsubsection{Phase 3 - Key distribution}
\label{subsubsec::IRK}
\vspace{-1mm}

The communication after Phase 2 will be encrypted with a {\em SessionKey} generated from LTK. BLE Encryption uses AES-CCM (Counter with CBC-MAC) and one {\em SessionKey} provides authentication and confidentiality. 

In Phase 3, the master and slave can exchange keys including the Identity Resolving Key (IRK) for device identity and privacy. 
BLE devices such as mobiles can be tracked if the MAC address is used in advertisement and in later communication. BLE addresses this privacy issue by IRK and a suite of protocols. 
IRK is used to generate resolvable private addresses in advertisement and communication. 
Only a device with privacy requirements needs to distribute its IRK and its real MAC address to its peer device. For example, if a mobile needs to protect its MAC address, it distributes its IRK and real MAC address to its peer device first. Then, the mobile uses this IRK to generate a resolvable private address for its packets and the peer device uses the mobile's IRK to resolve the private address.
If the peer device needs to protect its MAC address, it sends its own IRK and MAC address to the device for private address generation and resolution although such it is rarely used in practice.

\vspace{-2mm}
\subsection{Secure Connections Only Mode}
\label{subsec::SecureConnectionsOnly}
\vspace{-1mm}

For a slave device that provide services, the BLE specification defines the Secure Connections Only mode. % as authenticated LE secure connections pairing with encryption. 
This mode provides the highest BLE security level (Mode 1, Level 4 \cite{5.0-level4}), in which only the three secure pairing protocols, Passkey Entry, Numeric Comparison and secure OOB, can be used and the BLE Legacy is not allowed.
In this mode, if secure pairing is not used, the device shall send the Pairing Failed packet with the error code ``Authentication Requirements''. 

According to the BLE specification \cite{4.2sconly}, when a device is in the Secure Connections Only mode, ``The device shall only accept new outgoing and incoming service level connections for services that require Security Mode 1, Level 4 when the remote device supports LE Secure Connections and authenticated pairing is used.'' Here the service level connection refers to the application layer connection. 
It can be observed that BLE specifies the use of the Secure Connections Only mode for a slave that provides services, e.g. the blood pressure monitor in Figure \ref{fig:states}. However, the BLE specification does not explicitly define (or require) the Secure Connections Only mode for a master, e.g. the mobile in Figure \ref{fig:states}.

\vspace{-3mm}
\subsection{ATT Protocol}
\label{sec:gatt}
\vspace{-3mm}

The Attribute Protocol (ATT) is a server/client protocol with one BLE device as the server and the peer BLE device as the client. For example, the app on the mobile device is a client and the blood pressure monitor is a server in Figure \ref{fig:states}. A server maintains services in the format of attributes \cite{bluetooth2010bluetooth}. The client requests data from the server. 

An attribute has four properties: an attribute handle, a universally unique identifier (UUID), a value, and a set of permissions. The attribute handle is used to uniquely identify and access the attribute. If a client wants to read an attribute from a server, it issues a read request to the server with the handle. The UUID refers to the data type. The permission protects attributes on a device and specifies the security levels required in order to access attributes.
The permissions include read/write, encrypted read/write, authenticated read/write, and authorized read/write. A read/write attribute can be accessed with no restrictions.
An encrypted read/write attribute can only be accessed when pairing is applied and the link is encrypted. An authenticated read/write attribute can only be accessed when the link is encrypted with an authenticated-and-MITM-protection key%, i.e., secure pairing is used. 
An attribute with the authorized read/write permission can be accessed after authorization\footnote{The BLE specification does not explain its usage in detail.}.
It can be observed that the permission type is related to the pairing protocol type. If the attribute such as the keyboard input is sensitive, a high security level like authenticated read/write shall be used so that secure pairing protocols are used to counter eavesdropping and MITM attacks, and prevent keystroke leaking.

A Bluetooth profile specifies functionalities and features of each layer shown in Figure \ref{fig:overview} for a particular class of applications. A BLE device can implement the Generic Attribute Profile (GATT), which is built upon the ATT protocol, to exchange arbitrary data in the format of attributes with its peer devices. GATT assigns attributes into services and more details can be found in Appendix A.

%% As mentioned in specification, 
%Implementation of devices will determine how authorization occurs. Thus, in our paper, we will not take this type of attribute into consideration.}

\vspace{-2mm}
\section{Design Flaws in Android BLE Programming Framework}
\label{sec:pitfalls}
\vspace{-1mm}

As we mentioned earlier, the BLE specification defines the Secure Connections Only mode to ensure high security levels. 
With the Secure Connections Only mode enabled in a BLE device, the device will require secure pairing protocols to authenticate the initiator. But this mode is not enforced for the pairing initiator, e.g., a mobile, so the mobile is not required to authenticate the device. 
The lack of the mutual authentication with Secure Connections Only mode at both the device is in fact the cause of identified BLE security vulnerabilities. % and the mobile device in the BLE specification creates potential vulnerabilities. 

Interestingly, this insecure practice is strictly followed by modern BLE designs. 
Our analysis on the Android BLE programming framework proves that Android does not apply the Secure Connections Only mode. We examined the initiation, status management, error handling and bond management throughout the life cycle of a pairing process and identified four design flaws in Android as shown in Table \ref{tab:designflaws}. 
%We have performed extensive experiments and verified those design flaws. 

\begin{table}
%\vspace{-2mm}
\centering 
\caption{Summary of design flaws }
\label{tab:designflaws}
\begin{tabular}{|P{2cm}|P{5.6cm}|}
\toprule
{\bf Pairing stage} & {\bf Design flaw}    \\
\midrule
Initiation & No mechanism to specify a pairing protocol  \\
\hline
Status management & No mechanism to timely obtain the negotiated pairing protocol\\
\hline
Error handling & Mishandling pairing errors \\
\hline
Bond management &  No mechanism to remove a suspicious bond and start a new secure pairing process with a bonded device\\
\bottomrule
\end{tabular}
\end{table}

\smallskip\textbf{Design Flaw 1 - No mechanisms to specify a pairing protocol}. 
The function {\em createBond}() in Listing \ref{list:caculate} is the only function an Android app can use to start the pairing process with a peer BLE device. 
It does not accept any input parameter and the Android app cannot specify any pairing protocol even if it knows its peer BLE device's I/O capabilities. %capable pairing protocols.
The return value of this function, true or false, indicates if the pairing process has been successfully started. 
It can be observed from Listing \ref{list:caculate} that {\em createBond()} checks if the mobile device has an LTK in the device. If an LTK is available, {\em createBond()} returns false and will not re-pair with the peer device since the mobile device was paired with the device. 
It can also be observed that {\em createBond()} is an asynchronous call and does not wait for the pairing process to complete. Since Android OS services handle the pairing process, an Android app cannot pause or cancel the pairing process until the pairing is finished.
%That is, the pairing is handled by the Bluetooth stack 

\begin{lstlisting}[caption={The function \emph{createBond() } (Android 9.0)   },label={list:caculate},basicstyle=\footnotesize\ttfamily]
boolean createBond() {
    enforceCallingOrSelfPermission(BLUETOOTH_ADMIN_PERM,"Need BLUETOOTH ADMIN permission");
    //Checking the bluetooth admin permission 
    DeviceProperties deviceProp = mRemoteDevices.getDeviceProperties(device);
    
    //if already paired, return false
    if (deviceProp != null && deviceProp.getBondState() != BluetoothDevice.BOND_NONE) {
        return false;
    }
    ......
    //put a create bond message into the message processing queue
    Message msg = obtainMessage(BondStateMachine.CREATE_BOND);
    sendMessage(msg);
    return true;
}
\end{lstlisting}

\smallskip\textbf{Design Flaw 2 - No mechanisms to obtain the negotiated pairing protocol on time.}
Android provides asynchronous mechanisms for an app to know the status of a pairing process after pairing is completed. Through the intent \url{ACTION\_BOND\_STATE\_CHANGED}, the app knows pairing status including pairing in progress ({\em BOND\_BONDING}, pairing failure (\url{BOND\_NONE}), or pairing succeeded (\url{BOND\_BONDED}).  
Through the intent \url{ACTION\_PAIRING\_REQUEST}, the app knows either Passkey Entry or Numeric Comparison is adopted. 
By registering both intents \url{ACTION\_BOND\_STATE\_CHANGED} and \url{ACTION\_PAIRING\_REQUEST} (see more in Listing \ref{lst::determiningPairingMethod} in Appendix B),
an app knows the adopted pairing protocol, Passkey Entry, Numeric Comparison, Just Works or plaintext communication only after the pairing process is completed.  

The fact that an Android app knows the negotiated pairing protocol only after the pairing is completed breeds a security vulnerability of stealing a mobile's MAC address and IRK through the pairing process as shown in Section \ref{subsec:attackphone}. %about 
%In Section \ref{subsec:attackdevice}, we show an attacker can spoof a victim Android mobile device's peer BLE device and pair the spoofing BLE device with the Android mobile device via Just Works. 
%The attacker can then obtain the mobile device's MAC address and IRK through the pairing process before the app can tear down the BLE connection with the intent mechanism.
%Our evaluation in Section \ref{subsec:eva:appsec} also reveals that no BLE app uses the intent mechanism to identify the negotiated secure pairing protocol.
To defeat such attacks, we need to be able to tear down the connection at the time the pairing strategy is determined by exchanged I/O capabilities and before the MAC address and IRK are exchanged during the pairing process, rather than after the pairing process is completed. 

\smallskip\textbf{Design Flaw 3 - Android mishandles pairing errors.}
The Android Bluetooth service and stack do not memorize a negotiated pairing protocol. Further, Android does not provide APIs for apps to process pairing errors properly. Two possible pairing related errors in Android are introduced below.\looseness=-1

\smallskip{\textit{Pin or Key Missing (0x06)}}: When an Android mobile and its peer BLE device are paired, their communication link is encrypted with the negotiated keys including the LTK. 
If a peer BLE device's LTK is intentionally removed, the device will send an error code {\em 0x06} to the Android mobile during the connection process. 
But the Android mobile will not notify the user of this error. Instead, it will automatically communicate with the peer device in plaintext. 
Moreover, there are no APIs or mechanisms for an Android App to know the {\em 0x06} error ever occurred.

The incorrect processing of the {\em 0x06} error also creates a conflict between the bonding state and the link encryption state. When the {\em 0x06} error occurs, Android does not remove the corresponding LTK, which is supposed to encrypt the communication. The communication will be in plaintext while an Android app may use the Android reflection technique \cite{li2016reflection} to call a system level function {\em isEncrypted()} in order to check if the communication is in plaintext. 
However, the reflection technique is not allowed in the newer API since ``Using such methods or fields has a high risk of breaking your app'' \cite{restrictions-non-sdk}.

\smallskip{\textit{Insufficient Authentication (0x05)}}:
When an Android mobile tries to access an attribute with the {\em encrypted read/write} or {\em authenticated read/write} permission at the peer BLE device, the device will check whether the link is encrypted or a secure pairing protocol is used. If not, the peer device sends an error code {\em 0x05}, Insufficient Authentication, to the Android mobile.
After receiving the error code {\em 0x05}, the Android mobile's Bluetooth service starts a pairing process automatically for the app. The app can learn if the error occurs by checking the attribute access state code via a callback function {\em onCharacteristicRead}. The state code is {\em GATT\_INSUFFICIENT\_AUTHENTICATION} when the above error occurs. 
However, the app cannot stop the pairing process in this callback function. 
An attacker may spoof a paired device and utilize this {\em 0x05} error to start a pairing process with an Android mobile. This design flaw can be exploited to get the Android mobile's MAC address and IRK.

\smallskip\textbf{Design Flaw 4 - No mechanisms to remove a suspicious bond and to start a new secure pairing process with a bonded device.}
A third-party Android app cannot remove a bond from the mobile's list of bonded devices although the user can manually remove the bond with the system settings app. The function {\em removeBond()} can delete an LTK associated with the previously connected BLE device, it is a system level API and is not accessible by third party apps. 
Assuming that an Android app finds the insecure pairing Just Works is used, the app is able to break the connection. However, breaking the connection does not remove the bond, nor are the generated keys removed. 
The app can not start a new secure pairing process with a bonded device using \textit{createBond(.)} either.

\vspace{-2mm}
\section{Downgrade Attacks}
\label{sec:attacks}
\vspace{-1mm}

This section presents the threat model, attack overview, concrete attacks against mobiles and beyond mobiles.
%that downgrade BLE connections protected by secure pairing against Android mobiles. We then introduce attacks beyond mobiles, i.e., the attacks against the peer BLE devices of the mobiles.

\vspace{-2mm}
\subsection{Threat Model}
\label{sec:threatmodel}
\vspace{-1mm}

Our attacks against Android mobiles take the following assumptions. We believe that the assumptions can be easily fulfilled. (i) An attacker can obtain the same type of a victim device to explore its apps and communication protocols. For example, the attacker can purchase the same blood pressure monitor.
(ii) The attacker cannot physically access and unlock the mobile. 
%This is reasonable given that nowadays mobiles are sensitive gadgets and people normally tend them carefully. 
(iii) Our attacks do not need malicious apps installed on the mobile, one difference to many other attacks which require malicious apps for Bluetooth exploitation~\cite{NaveedZDWG14,sivakumaran2018attacks,xu2019badbluetooth,sivakumaran2019study}.
(iv) Before the attack, the Android mobile and its peer device are paired using secure pairing protocols such as the Passkey Entry and Numerical Comparison. %We make this assumption since it could be hard to attack the first time the device is configured.
%and it is more difficult to attack after the Android mobile and its peer device are paired with secure pairing. It will be trivial to attack an Android mobile app the first time it is used. An attacking device may perform the spoofing attack and pretend to be an legitimate device connecting to the mobile, which does not support the Secure Connections Only mode.
(v) The Android app is created using the official Android BLE programming framework.

\vspace{-2mm}
\subsection{Attack Overview}
\label{subsec::AttackOverview}
\vspace{-1mm}

Our attacks involve four parties, the sniffer, the blocker, a fake BLE device and a fake mobile. The Adafruit Bluefruit LE Sniffer \cite{Adafruit} is used to sniff BLE communication and collect basic information such as the device MAC address and the name from advertising packets and scan response packets. We use Texas Instruments (TI) CC2640 \cite{CC2640} development boards to simulate the blocker, the fake BLE device and the fake mobile.
%TI provides the Bluetooth Low Energy stack SDK and an Integrated development environment (IDE) called Code Composer Studio (CCS) for BLE application development.
%The development board offers peripherals such as Serial Peripheral Interface (SPI) and Universal Asynchronous Receiver Transmitter (UART) for hardware integration.

A blocker can launch a Denial of Service (DoS) attack and block a victim BLE device from connecting to a victim mobile so that a fake/spoofing device can connect to the victim mobile. 
We consider the following approaches.
(i) A jammer can be used to block the service of a victim device (not applied in this paper). (ii) A mobile acting as a master initiates a connection to its peer BLE device which acts as a slave. The number of connections to a slave is often limited. 
%Bluetooth 4.1 and early versions allow only one connection to a slave \cite{4.1,4.2}. 
BLE 4.2 and above allow multiple connections while the number of connections is up to the implementation  \cite{4.1,4.2}. For example, BLE 4.2 devices in our experiments allow one or three connections to the slave.
In case a slave allows multiple masters, multiple blockers can be used to connect to the victim BLE device and perform the DoS. Please note that connecting is different from pairing. A blocker can always connect to a victim peer device if no other smart device is connected to the victim device and even if the peer device requires secure pairing. Once enough blockers are connected to the victim device, no other smart device can connect to it.
(iii) A fake BLE device may increases its advertising frequency and connects to the victim mobile so that the victim device with the same MAC address cannot connect to the victim mobile. Our experiments in Section \ref{sec:eva:keyboard} have validated this approach and no blockers are needed.

The fake BLE device and fake mobile are full-fledged BLE devices and are also denoted as spoofing device and spoofing mobile. The BLE attack can be performed in the following steps. First, a fake device emulates a victim device.
The attacker can use a sniffer to obtain the MAC address and name of a BLE device. 
%Note that we did not find a peer BLE device of a mobile uses IRK, while a mobile often uses IRK to prevent itself from being tracked.  
A fake device is then configured to have the same MAC address and name as the victim BLE device. It can forge advertising and scan response packets that contain the same device name and service description as those of the victim device. The fake device can implement the same attributes of the victim device and manipulate the permissions of these attributes. 
Second, a fake mobile emulates a victim mobile. This requires that the fake mobile know the victim mobile's MAC address and IRK which is proved possible and will be demonstrated shortly.  

During an attack, the four parties will coordinate by the attacker to achieve a particular goal. For example, a blocker can be used to block a victim device so that the victim mobile will connect to the fake device. When the fake device is connected to the mobile, the attacker can change parameters of the BLE protocol such as the I/O capabilities. She can also intentionally create errors to poke the mobile.
When the fake device is connected to the victim mobile, the fake mobile can then connect to the victim device to perform a MITM attack, in which the fake device and fake mobile simulated by TI CC2640 can communicate through a UART port.

\vspace{-2mm}
\subsection{Downgrade Attacks against Mobiles}
\label{subsec:attackphone}
\vspace{-1mm}

We now show how an attacker can weaponize the design flaws in Section \ref{sec:pitfalls}, downgrade the pairing protocol established between a victim mobile and peer device, and perform more complicated attacks. 

\smallskip\textbf{False data injection via Design Flaw 3}:
To launch this attack, the fake device intentionally creates an error code {\em 0x06}. The communication between the Android mobile and the fake device is downgraded to plaintext as discussed in Design Flaw 3. We configure the permission of the attributes of the fake device as read/write so that the access to the attributes does not require any pairing. The fake device can then inject false data to the mobile. 
This attack cannot be easily detected since the Android mobile does not delete the original LTK. Therefore, even if the user checks the list of bonded devices at the Android mobile's system settings, the list will not show any aberrations. 

In the case of blood pressure monitoring, the false data injection attack may inject false blood pressure readings and misguide the doctors or nurses. Please note here we assume a scenario where BLE is adopted to connect medical equipment such as blood pressure monitors or x-ray imaging tools to data collecting equipment such as the tablet of a doctor or nurse. The purpose is to illustrate the potential threats of design flaws of Android. 

\smallskip\textbf{Spoofing attack on sensitive information via Design Flaw 3}:
In addition to the false data injection attack, the attacker can also utilize Design Flaw 3 and perform traffic analysis of the Android mobile. Through the use of Design Flaw 3, the attacker downgrades the communication to plaintext and the fake device can communicate with the Android mobile. 
Therefore, the fake device is positioned to receive any sensitive information from the Android mobile. 
We find that many IoT applications implement an application layer password mechanism for the user verification. When an authorized user inputs the password, the fake device can collect this password.

\smallskip\textbf{Stealing Android mobile's IRK and MAC address via Design Flaws 1, 2 and 3}: 
To protect the MAC address from leakage, an Android mobile with API 23 or above uses IRK introduced in Section \ref{subsubsec::IRK} by default~\cite{smartphoneuseIRK}. According to our experiments, the IRK is generated when the Android device is configured for the first time from the factory settings. It will not change until the device is reset to the factory settings. 
Any peer BLE device paired with the Android device will receive the same IRK and obtain the mobile's real MAC address.\looseness=-1

To obtain the IRK and MAC address of a victim Android mobile, the fake device needs to intentionally create a {\em 0x06} error message. The communication between the Android mobile and the fake device is then downgraded to plaintext.
The attacker also configures the permission of the attributes on the fake device as {\em encrypted read/write}. 
When the Android app tries to access these attributes, the fake device sends a ``0x05'' error message to the Android mobile, which starts a pairing process accordingly because of the Design Flaw 3. 
The fake device is configured to have no I/O capabilities so that the Android Mobile and the fake device pair with Just Works (See Design Flaws 1 and 2). Through the Step 10 in Figure \ref{fig:states}, the Android mobile distributes the IRK and MAC address to the attacker. With the IRK, the attacker can perform the private address resolution and trace the identity of the Android Mobile every time the mobile uses BLE.  We have confirmed this attack in our experiments. 
This attack defeats the purpose of IRK, which is used to prevent an Android mobile from being tracked.

\smallskip\textbf {Denial of Service (DoS) via Design Flaws 1,2,3 and 4}:
The goal of this DoS attack is to disrupt the communication between a victim mobile and its peer BLE device. According to the IRK stealing attack above, an attacker can first pair a fake device with a victim Android device using Just Works. This pairing process creates a new LTK for the mobile.
The attacker then turns off the fake device. The mobile then tries to communicate with the victim device. However, since the LTK on the mobile and the LTK on the victim device are now different, we find that Android cannot detect the two LTKs are different and the communication enters into a deadlock. As we mentioned in Design Flaw 4, there is no public API for an app to remove a bond on the mobile. The app cannot remove the bond or restart the pairing process. The deadlock can only be resolved by manually removing the bond in the Android system setting.

\vspace{-2mm}
\subsection{Attacks Beyond Mobiles}
\label{subsec:attackdevice}
\vspace{-1mm}

We have introduced attacks against Android mobiles. It is intuitive that those attacks will affect the peer BLE devices paired with the mobiles. We now discuss the attacks beyond mobiles.
The threat model for the attacks against peer BLE devices is different from the threat model for attacks against mobiles. In attacks against mobiles, we assume that the attacker cannot touch or unlock the victim mobiles. This is a reasonable assumption because people intend to tend their mobiles carefully. However, the attacker may have physical access to BLE devices in various scenarios. For example, IoT products such as smart lights may be placed outside. Few people physically lock away their BLE keyboards and attackers may press keys of those BLE keyboards.
Therefore, we consider the following two attack scenarios against peer BLE devices of mobiles. 

\textbf{Case 1. The attacker can physically access the victim BLE device}. 
Given a mobile cannot enforce secure pairing, an attacker with physical access to the peer BLE device can always launch an MITM attack even if the victim peer device uses the Secure Connections Only mode enforcing secure pairing. To deploy the MITM attack, a fake device connects to the victim mobile using the scheme in the false data injection attack. Since an attacker can touch and play with a victim peer device, she can always pair a fake mobile with the victim device, which may use any pairing protocol, even if the victim device enforces the Secure Connections Only mode. 
We show later even if the attackers can have physical access to peer BLE devices, once the secure connections only mode is enforced at both the Android device and peer device, our defense can still defeat those attackers. 

\textbf{Case 2. The attacker cannot physically access the device}. We also assume that before the attack, the Android mobile and its peer device are paired using secure pairing protocols such as Passkey Entry and Numerical Comparison.
Section \ref{sec:ble} shows that two secure measures can be adopted to protect sensitive data on a device, namely pairing and attribute permissions. Secure pairing protects the communication and attribute permissions limit access to attributes based on adopted pairing protocols.
We find that attribute permissions are often misused and the misused permissions will cause security issues.

\smallskip\textbf{Passive eavesdropping attack.}
This attack works when the victim device has only {\em read/write} attributes.  We assume that before the attack, the mobile pairs with the peer device that uses the Secure Connections Only mode. 
To launch this attack, the attacker first blocks the victim device. A fake device is then used to perform the {\em 0x06} error attack so that the communication between the fake device and the victim mobile is downgraded to plaintext. The fake device then goes offline and the blocker is turned off.
We find that the victim mobile then communicates with the victim peer BLE device in plaintext and will be able to access the the peer device's {\em read/write} attributes. Since the communication is in plaintext, the attacker can eavesdrop on the communication and retrieve sensitive information using a sniffer.
Similar to the false data injection attack, even if the user checks the bonded devices list at the mobile's system settings, no abnormalities will be observed. 

\smallskip\textbf{Bypassing the whitelist.}
A BLE device may use a whitelist of MAC address and IRK to allow connections only from an already paired mobile.
As discussed above, an attacker can steal a victim mobile's MAC address and IRK. Therefore a fake mobile with the same MAC address and IRK can bypass the whitelist mechanism and will be allowed to connect to the victim peer BLE device.
Even if the victim BLE device uses the Secure Connections Only mode, the fake mobile can still access read/write attributes of the BLE device after bypassing the whitelist.
If the permission of the attributes of the BLE device is encrypted read/write or authenticated read/write, the fake mobile has to pair with the peer device to access the attributes. If the BLE device enforces the Secure Connections Only mode or the attribute permission is authenticated read/write, the fake mobile has to perform secure pairing with the peer BLE device and may not be able to perform the attack. Recall that an authenticated read/write attribute requires secure pairing from the mobile.
The attack in Section \ref{sec:casestudy} presents a case study showing bypassing the whitelist leads to end-to-end MITM attacks against BLE keyboards .

\vspace{-2mm}
\section{Case Studies}
\label{sec:casestudy}
\vspace{-1mm}

In this section, we present case studies to demonstrate our attacks against Android and actual BLE products including BLE blood pressure monitors, smart lights, keyboards. Figure \ref{fig:light-test} in Appendix C shows 18 popular BLE products from various vendors that we have tested, including Logitech's K380 and K780 Keyboards \cite{k780}, Microsoft's Designer Keyboard \cite{Microsoft-k}, two blood pressure (BP) monitors from iHealth \cite{iHealth}, and QardioArm blood pressure monitor \cite{QardioArm}, TNG's Blood glucose meter \cite{TNG}, Flux light~\cite{Flux}, Magic Hue light \cite{MagicHue}, Magic Light \cite{MagicLight}, and MPOW light. 
%These products are  in their categories at Amazon.
Because of the page limit, please refer to Appendix C on the case study of blood pressure monitors demonstrating (i) the false data injection attack by exploiting Android mobiles (ii) passive eavesdropping attack against blood pressure monitors; the case study of smart lights demonstrating the spoofing attack stealing passwords with a fake light; the case study of popular Texas Instruments (TI)'s CC26XX development boards demonstrating the IRK and MAC address stealing attack and how we utilize it to control any BLE device using CC26XX, which implements the Secure Connections Only mode. All the attacks are launched without physical access to the Android Mobile and peer device. We focus on the case study of BLE Keyboards in this section. \looseness=-1

%\vspace{-1mm}
%\subsection{\bf BLE Keyboards} 
%\label{subsec::Keyboards}
%\vspace{-1mm}

\begin{figure} 
\centering
%\vspace{-3mm}
\includegraphics[width=0.8\columnwidth]{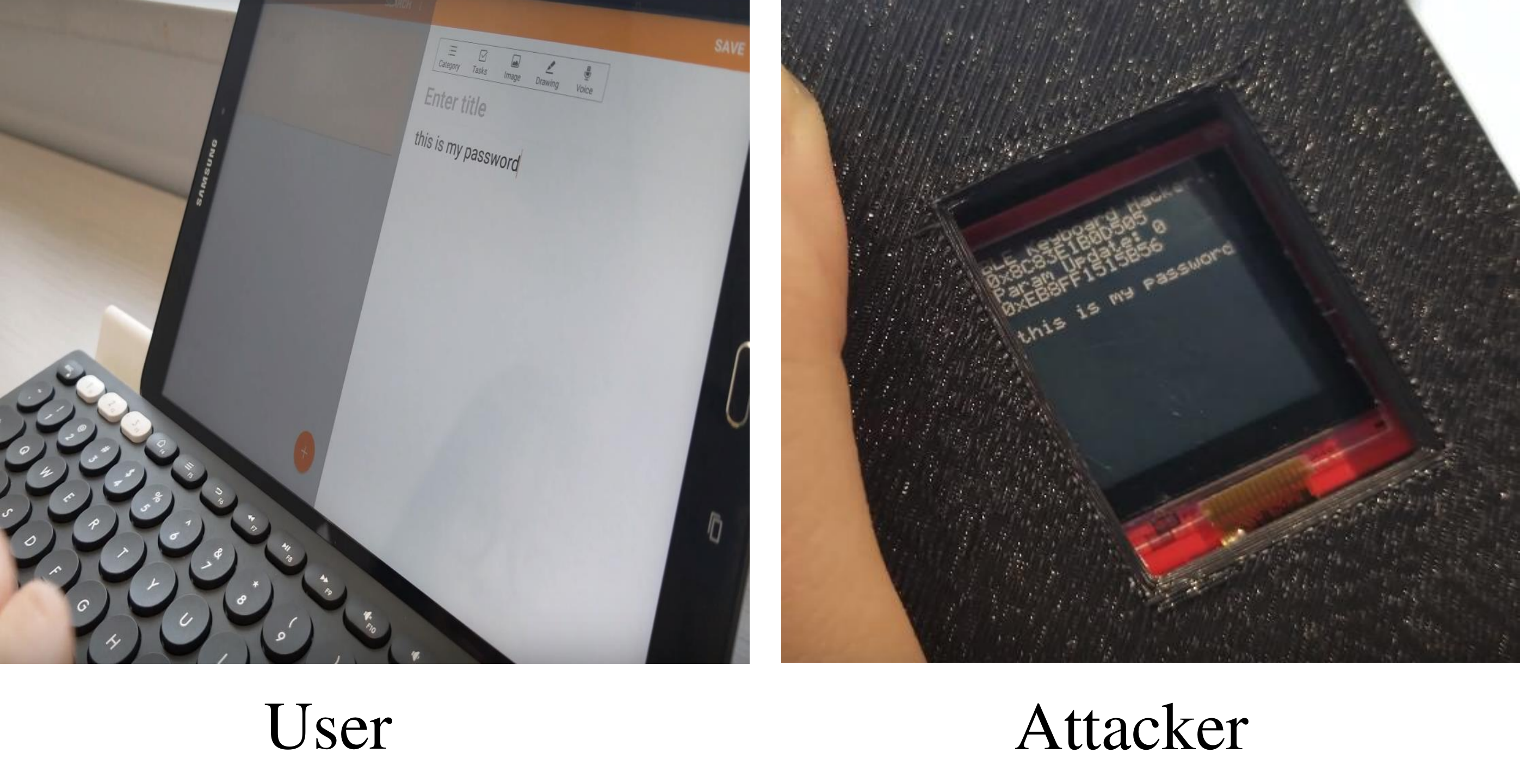}
%\vspace{-3mm}
\caption{\label{fig:keyboardhacker} MITM attack against a BLE keyboard from Logitech (An anonymous demo video is at \url{https://streamable.com/jgflo}. Please note there is no IP tracking. The attacking device is powered by a battery.)}
%\vspace{-3mm}
\end{figure}

In this BLE Keyboard case study, we exploit a BLE keyboard paired with an Android tablet to demonstrate (i) the attack to steal an Android mobile's IRK and MAC address and (ii) the attack to bypass the whitelist of the keyboard. Based on these two attacks, we also deploy the MITM attack against a victim keyboard. We have implemented the MITM attack against the BLE keyboard with two TI CC2640 development boards in a case as shown in Figure \ref{fig:keyboardhacker}. One board works as a fake tablet connecting to the victim BLE keyboard and the other as the fake BLE keyboard connecting to the victim tablet.

To steal the tablet's IRK and MAC address, we create a fake keyboard, which has the same MAC address and name as the victim keyboard.
The fake keyboard has a higher advertising frequency so that it has better chance to connect to the victim tablet instead of the victim keyboard.
We do not use a blocker here to block the victim BLE keyboard because a BLE keyboard often implements the whitelist mechanism and accepts only a previously paired mobile. The fake keyboard leverages the attack in Section \ref{subsec:attackphone} to obtain the IRK and MAC address of the victim tablet.

%We now discuss how to use the fake tablet with the stolen IRK and MAC address and a fake keyboard to deploy MITM attacks. 
The fake tablet with the stolen IRK and MAC address can bypass the whitelist mechanism of the victim keyboard. Since the current BLE Human Interface Device (HID) profile \cite{HID-1.0} does not enforce the Secure Connections Only mode and requires only encrypted (not authenticated) read/write permission for keyboard services, the fake tablet can {\bf remotely} pair with the keyboard with Just Works. The fake tablet and fake keyboard can then deploy a MITM attack.

%Given some lights only support {\em Just Works}, we also can deploy the MITM attack combing multiple downgrade attacks. For example, in the case of Lights from APPLights \cite{AppLights}, a fake device connects with the smartphone, and waits for the authentication data from the smartphone. A fake smartphone sets up another connection with the light. The fake device and fake smartphone can communicate with each other. Once the authentication data is received from the fake device, the fake smartphone forwards it to the light.
%Once the authentication process is completed, the fake smartphone can send commands to the light and fully control the light.

%We assume that the real device uses the Secure Connections Only mode and accepts pairing requests with the MITM flag setting to true. That is, the real device is supposed to accept only either Passkey Entry or Numeric Comparison for pairing. However, we find that if the real device uses TI's chip, by setting the MITM flag and declaring no I/O capabilities, the fake smartphone can pair with the real device using ``Just Works''.  

%TI’s SDK is faulty and cannot enforce this mode for secure pairing. Recall that The services offered by a BLE device can be protected by security permissions. For example,  However, TI’s SDK allows the authenticated read/write permission to be bypassed.

\vspace{-2mm}
\section{Countermeasures}
\label{sec:countermeasure}
\vspace{-1mm}

%We have studied security issues when BLE does not require the Secure Connections Only mode for connection initiators such as Android mobiles, which cannot enforce a secure pairing protocol.
In this section, we try to address the design flaws discussed in Section \ref{sec:pitfalls} and present solutions on how to enforce secure pairing within Android.
The Secure Connections Only mode shall be implemented as a configurable option for apps to defeat the presented attacks. %If the option is not used, BLE runs compatibly. 
%In this section, we propose defense techniques to fight the discovered attacks, which address the design flaws in android stack and provides a Secure connection only mode for apps to enforce the use of secure pairing.
We have implemented a prototype on Android 8 through the Android Open Source Project (AOSP)~\cite{AOSP}. Please note co-located attacks through malwares are addressed in \cite{NaveedZDWG14,sivakumaran2018attacks,xu2019badbluetooth,sivakumaran2019study} and are out of the scope of this work.
%Note that we advocate a Secure Connections Only mode as an option for Android BLE security while Android may keep its legacy BLE protocols to accommodate BLE applications that do not need strong security.
%\cite{NaveedZDWG14}\cite{sivakumaran2018attacks}

\vspace{-2mm}
\subsection{Overview}
\vspace{-1mm}

For a mission critical application, the app knows the peer device's I/O capabilities, which should support secure pairing.
With the Secure Connections Only mode enabled at the mobile, the user has to physically authenticate a BLE device.
If the negotiated pairing protocol between the mobile and its peer device is not the specified one, the communication shall be rejected and a critical security warning shall be directed to the user. 
Extra cost of a display or a keyboard and the usability issue of implementing secure pairing protocols may be associated with the proposed solutions. 
%However, this is a price that we have to pay in order to secure BLE.
\looseness=-1

\vspace{-2mm}
\subsection{Enforcing Secure Pairing}
\vspace{-1mm}

\smallskip{\bf Addressing Design Flaws 1 and 2: specifying a secure pairing protocol}. 
An Android mobile can enforce a secure pairing protocol after the mobile and peer device have determined the pairing protocol through the exchanged I/O capabilities, i.e., after Step 6 and before Step 7 in Figure \ref{fig:states}. 
If the negotiated pairing protocol is not the specified secure pairing protocol, Android should reject further actions and gives the user a security warning.
The Android BLE service and stack shall cache the specified secure pairing protocol in memory and save it in a configuration file on its nonvolatile storage if bonding is requested.

To address Design Flaw 1, an app can use our function {\em specifyPairing(.)} to store the specified pairing protocol in a configuration file \url{scm.conf} through the Java Native Interface (JNI). {\em specifyPairing(.)} is a system API. It can programmatically obtain the app's package name.
\url{scm.conf} is in the system folder \url{/data/misc/bluedroid/} and stores the app's package name and metadata including the specified pairing protocol. An app cannot manipulate metadata of another app.
When the pairing process starts, Android uses the system function \url{smp\_proc\_pair\_cmd()} to exchange the pairing features with the peer device. The bits in an integer \url{peer\_io\_caps} are used to indicate the peer device's I/O capabilities. Therefore,
\url{smp\_proc\_pair\_cmd()} can know the negotiated pairing protocol through announced I/O capabilities. In \url{smp\_proc\_pair\_cmd()}, we read the configuration file \url{scm.conf} and obtain the app's specified pairing protocol. If the specified pairing protocol and negotiated pairing protocol do not match, \url{smp\_proc\_pair\_cmd()} sends the error code \url{SMP\_PAIR\_AUTH\_FAIL} to the peer BLE device, halts the pairing process, breaks the connection and sends warning messages to the user.

Note that \url{smp\_proc\_pair\_cmd()} can obtain the negotiated pairing protocol at the earliest possible time. This also addresses Design Flaw 2. An app knows its specified pairing protocol will be enforced. If it cannot be enforced, the user will receive a security warning. 
Multiple apps on one mobile device could request pairing with the same BLE device although we did not find such use. If those apps require different pairing protocols, the one of the highest security shall be used.

%When launch an App for the first time, the app can use functions in \url{btif\_gatt\_client.cc} to configure \url{scm.conf} through the Java Native Interface (JNI).
%Each time when an App pair with a device, the app tells the Bluetooth stack its package-name for denoting the current running app. 

\smallskip{\bf Addressing Design Flaw 3: enforcing the specified paring protocol when errors occur}.
The {\em 0x06} error occurs because the fake device does not have the LTK. The {\em 0x05} error occurs because the BLE connection does not have the permission to access the attributes on the fake device. Android does not notify the user these errors and starts its own (vulnerable) pairing protocol.
We address the design flaw as follows. If there is any such pairing related error, the Android BLE service shall notify the user and ask the user whether to pair with the peer device. If the BLE connection has a specified pairing protocol and the user chooses to pair with the peer device, Android will enforce the specified pairing protocol and give the user a security warning if it cannot be enforced.

\smallskip{\bf Addressing Design Flaw 4: removing suspicious bond and starting a new secure pairing}.
An app shall be able to remove its own bonded devices whenever needed.
We make the system API {\em removeBond()} available to applications. 
{\em removeBond()} is redesigned so that {\bf an app can only remove its own bond}, not bonds of other apps. Therefore, a bond shall maintain metadata including the app's package name.
{\em removeBond()} will obtain the calling app's package name and can remove only its own bond. 
To deal with the rare case that multiple apps use the same device, we can use a mechanism similar to the Linux hard link \cite{hardlinks}. A counter is used to record the number of paired apps.  
%Only when all paired apps are removed will BLE remove the bond between the mobile device and the peer BLE device. 
%Every time an app tries to remove the bond and the counter is bigger than 1, the BLE stack notifies the user that other apps are paired with the device and if the user wants to remove all paired apps. If the user wants to remove all paired apps, all paired apps are removed and the bond is removed too. Otherwise, only the current app is removed and the counter decreases by one. In this way, the user can prevent a paired malicious app from removing the bond.
%Then, we provide a function {\em enforceBond()}, which uses the {\em removeBond()} as a building block. When this function called, the system first remove the bond created in previous time, then use the {\em createBond()} to pair with the intend device again. 
 
%\begin{figure}
%\centering
%\includegraphics[width=0.8\columnwidth]{Image/pairingdefense.pdf} 
%\caption{The architecture of our defense}
%\label{fig::defense}
%\end{figure} 
   
\vspace{-2mm}
\subsection{Security Analysis}
\label{subsec:secanalysis}
\vspace{-1mm}

We now discuss BLE pairing security if Android addresses the design flaws and enforces secure pairing, and the peer BLE device also enforces secure pairing. Under the assumption that an attacker cannot physically access the mobile or peer BLE device, the attacks in Section \ref{sec:attacks} will fail since secure pairing requires the attacker (fake mobile) see the victim device and the mobile see the attacker (fake device). 

\begin{figure} 
\centering
\includegraphics[width=\columnwidth]{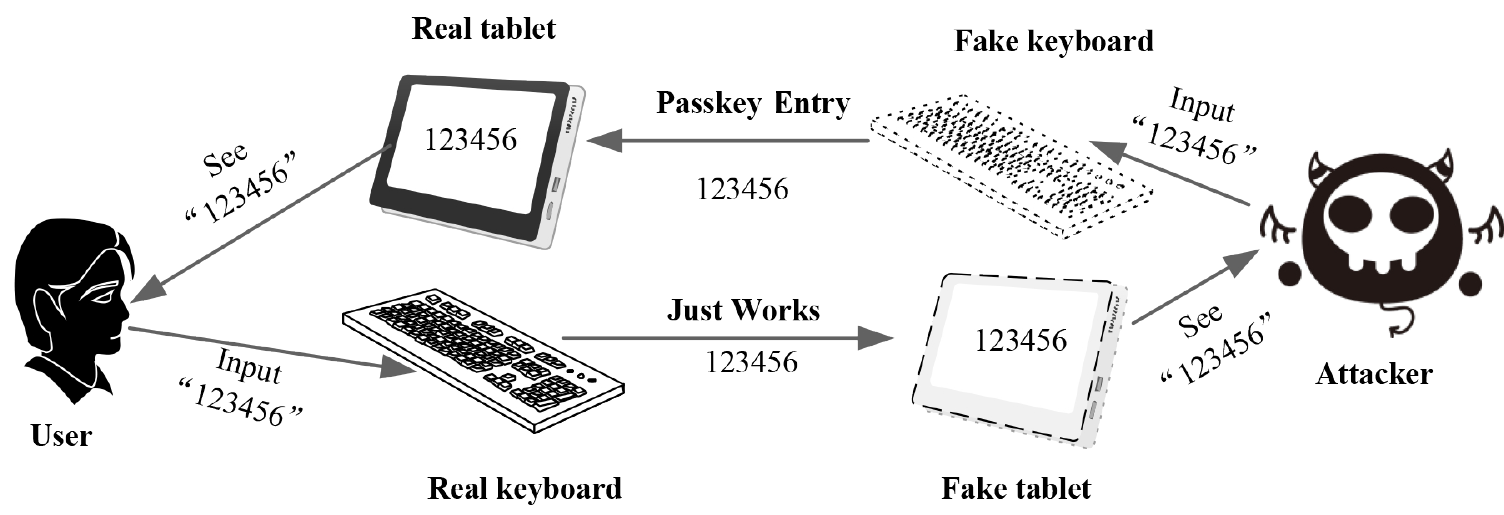}
%\vspace{-2mm}
\caption{The work-flow of attacking the Keyboard with Passkey Entry enforced}
\label{fig::K830}
%\vspace{-2mm}
\end{figure}

%\subsubsection{Passkey Entry vs. Numerical Comparison}
As discussed in Section \ref{subsec:attackdevice}, 
the assumption of no physical access is not always true. Although we can assume the attacker cannot physically access a mobile, we cannot assume the attacker cannot always access a BLE device such as a BLE keyboard. 
if an attacker can physically touch a BLE keyboard that uses the Passkey Entry pairing protocol, even if both the keyboard and mobile enforces passkey entry, the attacker can still perform the MITM attack as follows. 
The Passkey Entry pairing protocol is secure only if the attacker cannot obtain the passkey. However, the BLE keyboard is a human input device, which sends whatever keystrokes to a mobile device as long as the mobile device is paired with the keyboards. As shown in Figure \ref{fig::K830}, if the attacker can physically access the keyboard, she can pair the fake mobile with the keyboard by entering a chosen passkey when the user is away from the device. The fake keyboard later pretends to be the real one and starts a pairing process with the user's victim mobile. The victim mobile enforces Passkey Entry and requires the user enter a passkey displayed on the victim mobile. 
However, when the user enters the passkey on the victim keyboard, the fake mobile receives the user entered passkey. The fake mobile then sends the passkey to the fake keyboard, which can then use the passkey to connect to the victim mobile. The attacker can now perform the MITM attack.

The MITM attack above will fail when the victim mobile and keyboard enforce the Numeric Comparison pairing protocol even under the assumption that the attacker can physically access the keyboard.
To implement Numerical Comparison, the keyboard must have a display.
The attacker's fake mobile can still be paired with the victim keyboard because of the assumption of physical access. 
However, when the user pairs the victim keyboard with the victim mobile, the user has to compare the two numbers displayed on the victim keyboard and the victim mobile. With the underlying Numerical Comparison protocol, if the attacker performs the MITM attack with a fake mobile and a fake keyboard in the middle, the two numbers on the victim keyboard and the victim mobile will be different. 
%Please refer to the Numerical Comparison protocol in Section \ref{subsubsec:pairprotocols}. 
The MITM attack will be detected and fail.\looseness=-1

Based on the analysis above, it can be observed that under the assumption that the attacker can physically access the keyboard, Numerical Comparison is more secure than Passkey Entry. 
When we enforce secure pairing, Numerical Comparison provides the strongest pairing security. 
The BLE specification treats Passkey Entry and Numeric Comparison equally and these two secure pairing protocols have the same security level - authenticated-and-MITM-protection. In the specification, if either of the two protocols is applied, the connection is considered as authenticated. This term is not accurate based on our analyses.

\vspace{-2mm}
\section{Attack and Defense Evaluations} 
\label{sec:eva}
%\vspace{-1mm}

In this section, we first evaluate our attacks, and then evaluate our defense measures and its feasibility. 
%We present our preliminary study of securing pairing on iOS at last.

\vspace{-2mm}
\subsection{Android OSes}
\vspace{-1mm}

We tested all design flaws on the mainstream Android versions, from 7.0 to 9.0 and find that all our attacks can be performed with no adjustments (see Table~\ref{tab:phone-test}).
Recall that a fake device may use the {\em 0x05} error in Section \ref{sec:pitfalls} to stealthily 
pair with the victim mobile through Just Works. This approach works under all versions of Android we tested.
On Android 7.0, a fake device can also send a security request as introduced in Section \ref{sec:pairfeature} to stealthily pair with the victim mobile while the security request on higher versions of Android will raise a pairing request dialog window asking the user for permission. Such a dialog Window may alert the user.

\begin{table} 
\centering 
\vspace{-2mm}
\caption{Summary of Android mobiles under testing }
\label{tab:phone-test}
\begin{tabular}{|c|c|c|}
\toprule
Brand &  OS    \\
\midrule
Samsung Galaxy S8+  & Samsung Official Android 7.0  \\
\hline
Google Pixel 2 & AOSP Android 8.0  \\
\hline
Samsung Tablet & Samsung Official Android 8.1  \\
\hline
Samsung Note 8  &  Samsung Official Android 8.1  \\
\hline
Google Pixel 2  &  AOSP Android 9.0  \\
\bottomrule
\end{tabular}
\vspace{-2mm}
\end{table}

\begin{table} 
\centering 
\vspace{-2mm}
\caption{BLE apps at Androzoo}
\label{tab:app-intents}
\begin{tabular}{|P{2.2in}|P{0.7in}|}
\toprule
{\bf Apps} & {\bf Quantity}  \\
\midrule
BLE apps & 3501 \\
\hline
BLE apps using {\em createBond()} & 176 \\
\hline
BLE Apps using intents for pairing status & 29 \\
\bottomrule 
\end{tabular}
%\vspace{-3mm}
\end{table}

%\vspace{-2mm}
\subsection{BLE App Security}
\label{subsec:eva:appsec}
\vspace{-1mm}

To evaluate the security of BLE apps, we collected 3501 Android apps from the Androzoo database \cite{allix2016androzoo}, which actively collects apps from all major Android app stores including Google Play. All 3501 BLE apps we evaluated are subject to our downgrade attacks. 
Specifically, we build a tool based on the public framework {\em soot} \cite{vallee2000optimizing} to scan Android apps and examine how they perform pairing related functionalities. 
%At the time of writing, Androzoo has 8,443,373 apks.
Androzoo provides a set of RESTful APIs \cite{RESTfulAPI} to query apps of interest. 
To find BLE related apps, we set the query condition as apps that use {\em connectGatt()}, which a BLE app must use to connect to a peer BLE device.
%This function is used to setup a connection of BLE.
Among the collected 3501 BLE apps, only 176 apps use the {\em createBond()} to explicitly start a pairing process and 29 apps use the intent mechanism (see Table \ref{tab:app-intents}). 

We manually analyze the 29 apps that use the intent \url{ACTION\_BOND\_STATE\_CHANGED} or \url{ACTION\_PAIRING\_REQUEST}. These intents are used for various reasons. Some apps register the intents to determine if the mobile is bonded with the intended device so that it can transmit data. This cannot deter our downgrade attacks since the victim mobile and the victim peer device are bonded under our attacks.
We also find that many apps using the programming frameworks in a faulty way.
Some apps use \url{ACTION\_PAIRING\_REQUEST} to determine if Passkey Entry or Numeric Comparison is used with the consideration of usability rather than security.
These apps automatically input a fixed passkey for the Passkey Entry pairing protocol or ``click'' the confirmation button for Numeric Comparison when its peer device adopts Numeric Comparison. 
These strategies make Passkey Entry and Numeric Comparison useless since the user is not involved.

%\vspace{-2mm}
\subsection{Keyboard Connection Competition}
\label{sec:eva:keyboard}
\vspace{-1mm}

\begin{figure} 
\centering
\vspace{-2mm}
\includegraphics[width=0.8\columnwidth]{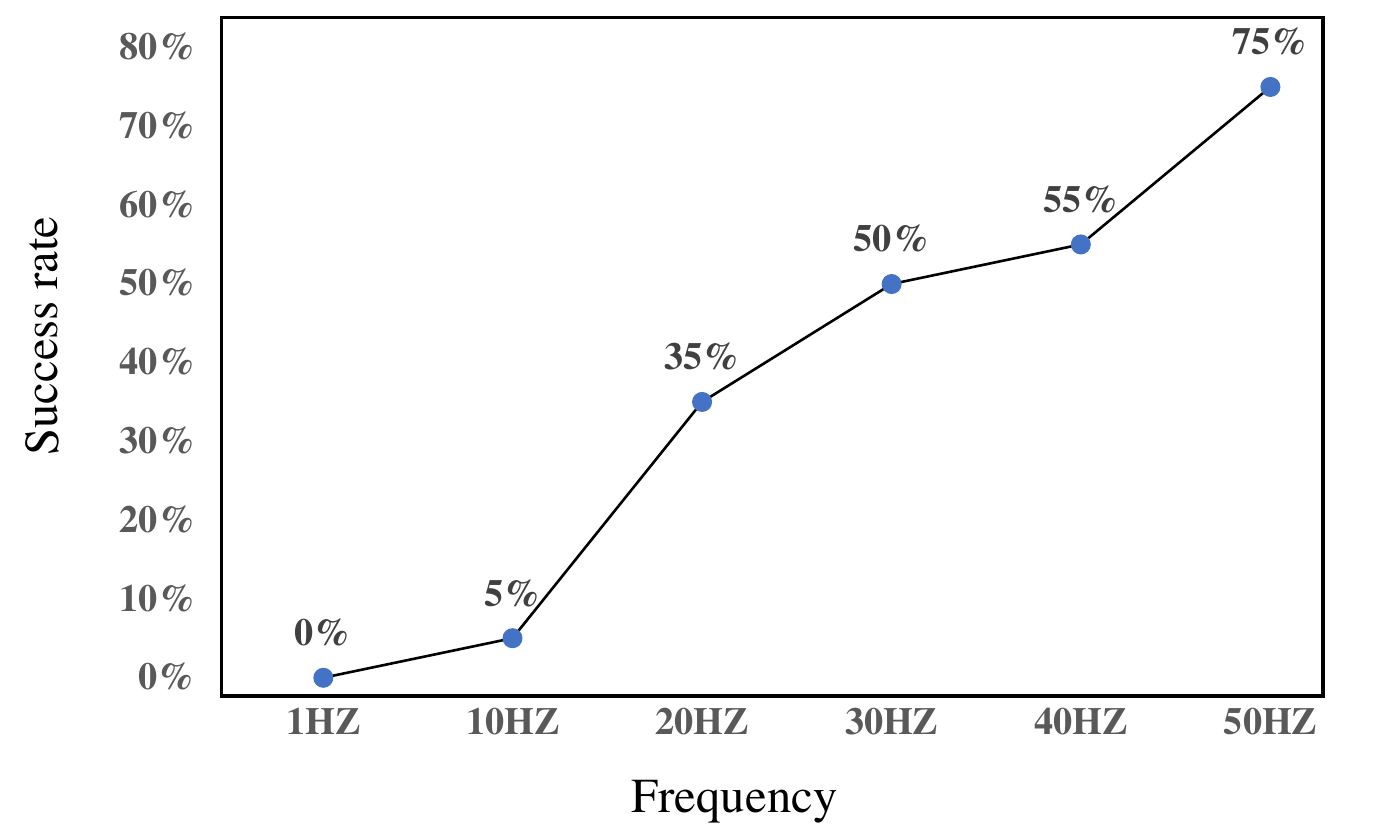} 
%\vspace{-2mm}
\caption{Impact of advertising frequency}
\label{fig:adv}
%\vspace{-2mm}
\end{figure}

As discussed in Section \ref{sec:casestudy}, when both the victim keyboard and a fake keyboard try to connect to the victim mobile, the one with a higher advertising frequency has a better chance. We now present the impact of the advertising frequency on the success rate of the fake keyboard connecting to the victim mobile.
In our experiments, the victim keyboard is put close to an Android mobile as in a normal use scenario, while the fake keyboard is 10 meters away from the keyboard.
For each advertising frequency, we perform the connection competition game 20 times. The success rate is the number of successful connections by our fake keyboard over 20. 
Figure \ref{fig:adv} shows the success rate versus the advertising frequency. The success rate is 50\% when the advertising frequency of the fake keyboard is 30HZ. The BLE specification sets the highest advertising frequency as 50 HZ, at which the success rate by the fake keyboard is 75\%. We use CC2640 for the fake keyboard, which does not work when the advertising frequency is beyond 50HZ.
%In general, we can see that the attack can be deployed for the most time.

%In Table \ref{tab:phone-test},  $\times$ refers to unsuccessful attacks. $\foobar$ means that a pairing request window is displayed and the attack succeeds if a user accepts the pairing request. All the attacks are successful against Android. iOS can defend against the downgrade attack 2 in Table \ref{tab:phone-test}, since iOS always starts the encryption as long as an LTK is present.For the downgrade attack 3, iOS shows a pairing dialog box. Attacks 1 and 4 work against iOS with no change. 
 
\vspace{-2mm}
\subsection{Countermeasure Evaluations}
\vspace{-1mm}

We have implemented our defense techniques on a Google Pixel 2 mobile through the AOSP. We launched all our attacks and confirmed that they failed under the patched Android system. 
For example, in the case of the BLE keyboard, when Numerical Comparison is enforced, the user finds that the two numbers displayed on the victim mobile and keyboard (emulated by a CC2640) are different when the MITM attack is deployed. The user shall reject the pairing and investigate the possibility of attacks. 
\looseness=-1 
%For example, in the case of the blood pressure monitor in Section \ref{subsecbp}, when the 0x06 error occurs, Android will enforce the Passkey Entry protocol specified by the app. At this time, the attacker has no chance to inject false data to the mobile app. In the case of a keyboard, the HID profile, which can be considered as an app, can specify
%Numeric Comparison when the pairing process starts for the
%strongest security. Thus, the downgrade attacks have been mitigated by using our countermeasure. 

\begin{figure} 
\centering
\vspace{-2mm}
\includegraphics[width=0.75\columnwidth]{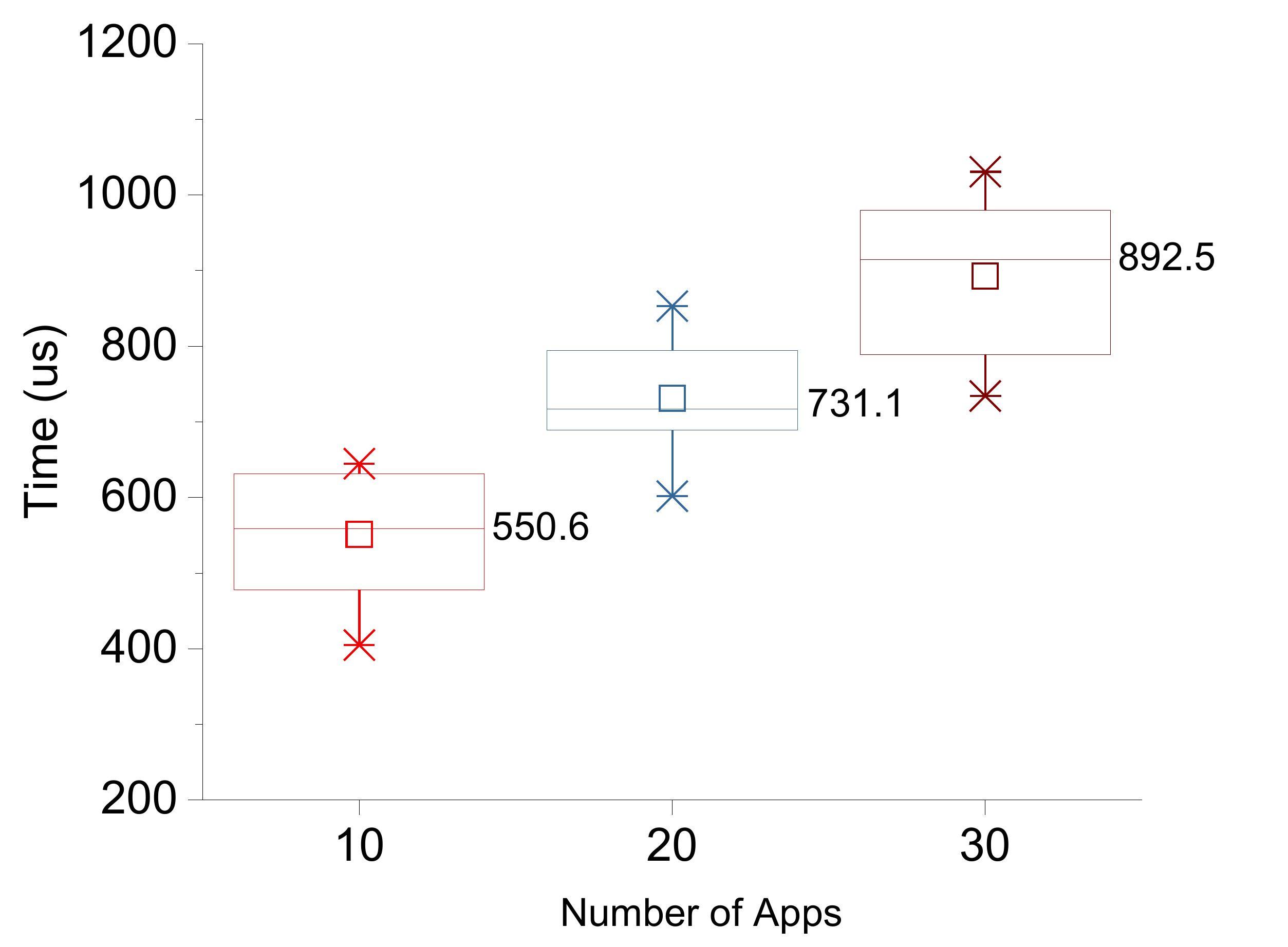} 
\vspace{-2mm}
\caption{Defense performance}
\vspace{-2mm}
\label{fig:defense:eva}
\end{figure}

We also evaluate the performance of our secure pairing strategy, i.e. the overhead caused by the query of the configuration file {\em scm.conf} for a specific app's metadata such as the specified pairing protocol. We tested three cases: 10, 20 and 30 BLE apps using our defense mechanisms on the security enhanced Android mobile. The app of interest is always set as the last one in {\em scm.conf}. That is, we consider the worst case of time needed to find the metadata of the app of interest.
We run the test for each case 10 times and derive the average time.
As shown in Figure \ref{fig:defense:eva}, the average delay is from 550.6$\mu$s to 892.5$\mu$s and is feasible for typical use of BLE apps in a mobile \cite{howmanyapps}.
 
\section{Discussions}
\label{sec:discuss}
\vspace{-1mm}

\subsection{Securing Pairing on iOS}

%{\bf Downgrade attacks against iOS}. 
According to the iOS developer guidelines \cite{IOS-pairing}, iOS does not provide a way of enforcing secure pairing protocols either. 
We also performed all developed attacks against iOS.
The false data injection and spoofing attack for sensitive information are the same against iOS. That is, when the {\em 0x06} error occurs, the iOS will not notify the app and user and run the vulnerable protocols similar to Android. 
The downgrade attack stealing the IRK and mobile's address does not work since iOS does not respond to the {\em 0x05} error and does not notify the user the error either. This could be a problem of iOS since it misses the error processing.
An attacker may steal the IRK and MAC address in another way. We find some apps try to pair with any device of the same model and manufacturer automatically. Since iOS cannot enforce secure pairing, a fake device emulating a device of the same model and manufacturer can use Just Works to pair with iOS. In this case, iOS displays an innocuous dialog box and shows a message such as ``Device X would like to pair with your iPhone'', where Device X is the peer device's name. If the user clicks the ``Pair'' button, the attacker obtains the IRK and MAC address of the iPhone.
The DoS attack does not work against iOS since iOS does not respond to the 0x05 error.

\subsection{Other Solutions to BLE Security}

There are other ways to secure BLE communication. Many chips such as TI's CC3220 \cite{CC3220} now support crypto accelerators or can be equipped with a low-cost co-processor such as ATECC608A which only costs (\$0.55) \cite{ATECC608A}. Each device is equipped with a pair of (public key, private key). The private key is stored in the secure storage while the public key is published, for example, as a tamper resistant QR code label on the device's exterior. A mobile can now get the device's public key and use public key cryptography such as Elliptic-curve cryptography (ECC) to implement a SSL/TLS-like protocol at the application layer to secure its communication with the peer device, using BLE as only a wireless media. The disadvantage of this approach is it requires extra hardware support.

Another approach is we assume there is no attack the first time the user configures the mobile and device, which can share a secret to protect later communication. The disadvantage of this approach is the assumption may not be true. Our solution of enforcing secure pairing at both the mobile and device can address these issues while defeating other attacks such as co-located attacks requires extra remedies and is addressed in \cite{NaveedZDWG14,sivakumaran2018attacks,sivakumaran2019study}.

\vspace{-2mm}
\section{Related Work} 
\label{sec:relatedwork}

%In this section, we review most related work on Bluetooth vulnerabilities and attacks, and focus on BLE.
\smallskip{\bf Vulnerabilities in Bluetooth:} 
%Before the Simple Secure Pairing (SSP) was introduced by Bluetooth, Bluetooth suffered from the brute force attack \cite{becker2007bluetooth,shaked2005cracking}, eavesdropping \cite{spill2007bluesniff}, MITM attack \cite{kugler2003man} and relay attack \cite{levi2004relay}. 
Bluetooth before the Simple Secure Pairing (SSP) is not secure \cite{becker2007bluetooth,spill2007bluesniff, shaked2005cracking,levi2004relay,kugler2003man} and is out of scope of this paper. The Simple Secure Pairing is also vulnerable. For example, Haataja et al. \cite{haataja2010two,hypponen2007nino} proposed MITM attacks against SSP of Bluetooth Classic (versions 2.1 and 3.0) in 2010. They assumed that the victim devices use only I/O capabilities to determine the pairing strategy and the attacking devices can pair with victim devices using Just Works. The latest BLE introduces the Secure Connections Only mode in order to defeat those attacks. Our work focuses on the Secure Connections Only mode in Android. 

Mike Ryan \cite{ryan2013bluetooth} built a BLE sniffer over Ubertooth and demonstrated that the Passkey Entry pairing protocol for LE legacy connections is not secure. His tool {\em crackle} can crack such connections and target BLE 4.0 and 4.1. Our paper addresses the latest BLE 4.2 and 5.x, which are considered secure against his attacks. The work by Rosa \cite{rosa2013bypassing} is similar to Mike Ryan's work.
Working in the area of telemedicine, Zegeye et al. cracked the BLE temporary key used in the pairing process by using a brute-force attack \cite{zegeye2015exploiting}, which also extends the attack in \cite{ryan2013bluetooth}.
Dazhi Sun et al. \cite{sun2018man} proposed a method that can break Passkey Entry when the passkey is reused. The similar problem was also discussed in \cite{sivakumaran2018low}. However, reusing a passkey is not recommended in BLE, which requires a random passkey shall be used in each pairing session with Passkey Entry. We assume a random passkey in this paper.

\smallskip{\bf Bluetooth attacks on mobiles:} 
Jasek et al. \cite{jasek2016gattacking} studied possible attacks between a Bluetooth smart device and its mobile app. However, they study BLE 4.0 and 4.1, which do not have the Secure Connections Only mode for BLE. They attacked Passkey Entry with Mike Ryan's approach \cite{ryan2013bluetooth}. 
Many works reverse engineer particular products \cite{bouhenguel2008bluetooth,cyr2014security,zhang2017security,cusack2017assessment,das2016uncovering} and exploit the faulty app protocols while we focus on the operating system level and programming framework issues. For example, Britt Cyr et al. performed a security analysis of wearable fitness devices~\cite{cyr2014security}. They reverse engineered the devices, BLE communication traffic, and app, and used Mike Ryan's attacks against pairing.
Zhang et al. analyzed the commands from four popular smart wristbands by sniffing packets without reverse engineering the apps \cite{zhang2017security}, and presented replay and MITM attacks against those particular wristbands. 
Brian Cusack et al. investigated vulnerabilities of BLE in mainstream wearable devices~\cite{cusack2017assessment}, and showed that most of the wearables are subject to privacy disclosure. BlueBorne \cite{BlueBorne} explored faulty BLE implementations. our attacks are not based on those issues.
William et al. \cite{oliff2017evaluating} and Melamed et al. \cite{melamed2018active} studied the spoofing attack and MITM attack between a Bluetooth smart device and its mobile app. They presented the software based and hardware based attacks, but did not address how to attack two paired devices with a secure pairing protocol. Fawaz et al. \cite{fawaz2016protecting} collected and analyzed the advertisement packets from 214 BLE devices. They found that the poor design and implementation of BLE advertisements may lead to privacy leaks. We address pairing security in this paper.
Muhammad Naveed et al. \cite{NaveedZDWG14}, Xu et al. \cite{xu2019badbluetooth} and Sivakumaran et al. \cite{sivakumaran2018attacks,sivakumaran2019study} also addressed Bluetooth security but not on the pairing process.

\vspace{-3mm}
\section{Conclusion}
\label{sec:conclusion}

BLE 4.2 and 5.x have a Secure Connections Only mode to enforce secure pairing such as Passkey Entry and Numerical Comparison for BLE devices. However, the BLE specification does not explicitly require an initiating device such as a mobile to support the Secure Connections Only mode. This creates potential security vulnerabilities against both mobiles and their peer BLE devices. In this paper, we systematically investigate Android's BLE programming framework and present four design flaws of its pairing protocol. We then present a suite of downgrade attacks and case studies exploiting these design flaws against Android mobiles and peer BLE devices.
%Through the case study of BLE keyboards, we also find that Passkey Entry is less secure than Numerical Comparison.
To defend against these attacks, we patch the discovered design flaws and enforce secure pairing of Android. %If the negotiated pairing protocol between the mobile and peer BLE device is not the specified one, further communication is rejected and a security warning is sent to the user. 
%If errors occur during communication between two bonded devices and a secure pairing protocol is specified, repairing is started and the specified pairing protocol is enforced. 
We have also performed extensive evaluations to validate the discovered attacks and proposed defense mechanisms. In our future work, we will implement and evaluate hardware-supported solutions for BLE security. We will also explore application layer security solutions for BLE devices.

\bibliographystyle{IEEEtran}
\bibliography{USENIX2019}
%\clearpage 
 
\section*{Appendix A - BLE Profiles}
\setcounter{subsection}{0}

A Bluetooth profile specifies functionalities and features of all layers in Figure \ref{fig:overview} for a particular class of applications. A profile can be considered as a specific application. For example, the Human Interface Device Profile (HID) defines rules that allow a HID device, such as a keyboard, using Bluetooth to accept inputs from humans and shows the output to humans. A profile may contain other profiles and protocols as its building blocks to implement functionalities. The Generic Access Profile (GAP) defines the basic requirements of a Bluetooth device and all Bluetooth devices implement GAP. For example, GAP performs advertising and scanning. 

A smart device can implement the Generic Attribute Profile (GATT), which is built upon the ATT protocol, to exchange arbitrary data in the format of attributes with its peer devices. 
GATT organizes attributes into services.
A service contains zero or more characteristics, which are also attributes and user data containers. 
A characteristic contains zero or more descriptors, which provide more metadata.
A primary service provides the primary functionality of the device. A secondary service can work as a building block and should be included in the primary service. 
 
\section*{Appendix B - Using Current Android Mechanisms to Detect Pairing protocol}
\setcounter{subsection}{0}
 
 %\begin{figure} 
%\includegraphics[width=\columnwidth]{Image/lifecycle.pdf}
 
%\caption{\label{fig:lifecycle} Overview of Bluetooth Low Energy Programming Framework}
 
%\end{figure}
 
In Listing \ref{lst::determiningPairingMethod}, by registering intent \url{ACTION\_PAIRING\_REQUEST} and \url{ACTION\_BOND\_STATE\_CHANGED}, an Android app can asynchronously know the adopted pairing method, Passkey Entry, Numeric Comparison, Just Works or plaintext communication. 

\begin{lstlisting}[caption={Android determining pairing method after bonded}, label={lst::determiningPairingMethod},basicstyle=\footnotesize\ttfamily]
boolean numericcomparison=false;
boolean passkey=false;
boolean justworks=false;
boolean plaintext=true;

// Activity starts; register intents
public void OnCreate(){ 
  IntentFilter pairingRequestFilter = new IntentFilter();
  pairingRequestFilter.addAction(BluetoothDevice.ACTION_BOND_STATE_CHANGED);
  pairingRequestFilter.addAction(BluetoothDevice.ACTION_PAIRING_REQUEST);
  registerReceiver(mPairingRequestRecevier, pairingRequestFilter);
    ....
  }

    ....
  //Once connected call createBond()
  device.createBond();

// Process intents and determine pairing method
public void onReceive(Context context, Intent intent) { 
  if (BluetoothDevice.ACTION_PAIRING_REQUEST.equals(intent.getAction())){ // either numeric comparison or passkey is used
    int pairingtype = intent.getIntExtra(BluetoothDevice.EXTRA_PAIRING_VARIANT, BluetoothDevice.ERROR);
    
    if(pairingtype==BluetoothDevice.PAIRING_VARIANT_PASSKEY_CONFIRMATION){
      numericcomparison=true;
      plaintext=false;
    }
    if(pairingtype==BluetoothDevice.PAIRING_VARIANT_PIN){
      Passkey=true;
      plaintext=false;
    }
  }
  	
  if (BluetoothDevice.ACTION_BOND_STATE_CHANGED.equals(intent.getAction())) { // Bonding, bonded, or bonding none (failure)?
    int bondstate = intent.getIntExtra(BluetoothDevice.EXTRA_BOND_STATE, BluetoothDevice.ERROR);
            
	if(bondstate==BluetoothDevice.BOND_BONDED){			   
	  if(!numericcomparison || !passkey){
	    justworks=true;
      plaintext=false;
	  }
	}
 }          
}
\end{lstlisting} 
   
\section*{Appendix C - Case Studies}
\setcounter{subsection}{0}

\textbf{Blood Pressure Monitors}: In this case study, we demonstrate that (i) the false data injection attack by exploiting Android mobiles; and (ii) passive eavesdropping attack against blood pressure monitors. Blood pressure monitors in our experiments may use a secure pairing method to pair with an Android mobile. For example, the iBalance Blood pressure monitor has a display and a button, and uses the Passkey Entry to pair with a mobile. The app collects readings from the monitor. 
To attack Blood Pressure monitors, we reverse engineered the app to understand its application layer protocol. We are able to deploy the downgrade attack against the mobile for false data injection. 
Particularly, a fake monitor is created using a TI CC2640 development board, which sends false blood pressure measurements to the mobile.
The attributes of all the blood pressure monitors in our experiments are configured as read/write. Therefore, they are subject to our passive eavesdropping attack.

\begin{figure}  
\centering
\includegraphics[width=0.9\columnwidth]{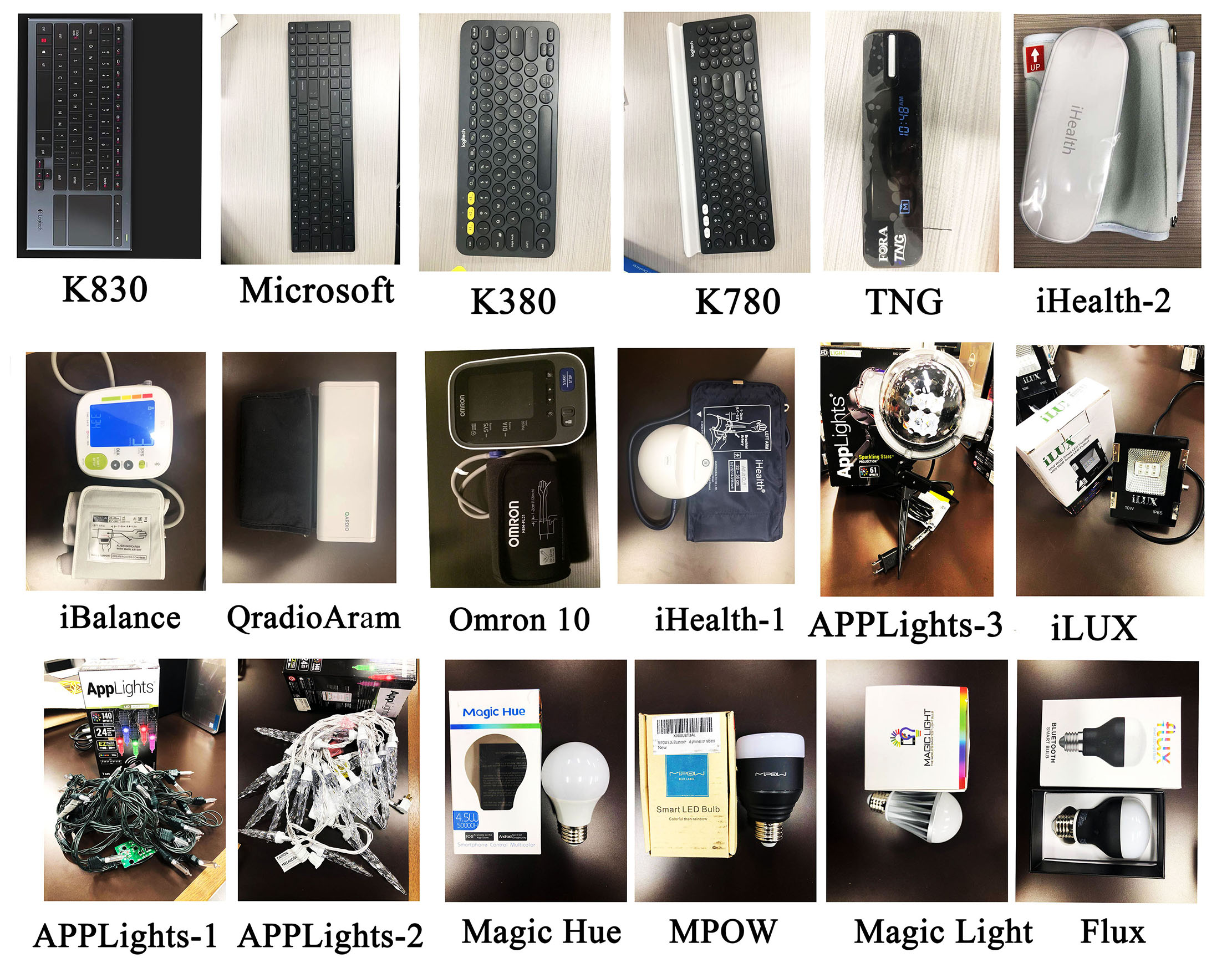}
\caption{\label{fig:light-test} Tested devices in our experiments}
\end{figure}

\textbf{Smart Lights}: 
In this case study, we demonstrate that (i) the spoofing attack stealing passwords by a fake light; and (ii) DoS attack.
Recall that a BLE app may use passwords to implement authorization and limit who can use the devices. In our experiments, we find the Flux smart light uses the Passkey Entry pairing protocol. We assume the app uses secure pairing to securely set up the password on the light. To steal the password, an attacker can use a blocker to block the service of the light. A fake light can then be positioned to connect to the Android mobile when the user opens the app to use the light. The app will send the password to the fake light controlled by the attacker, who can then control the BLE light. 

To deploy the DoS attack, the fake light with the victim light's MAC address uses Just Works and pairs with the victim mobile so that a new LTK will be generated for the victim mobile. The fake light then goes offline. The communication between the mobile and the real light then fails since the LTKs on the two sides are different. As discussed in Section \ref{subsec:attackphone}, the mobile can not access the service provided by the real light until the user manually removes the bond with the Android system setting app.

\textbf{CC26XX Development Boards from Texas Instruments}: We now use the case study to demonstrate the impact of leaking a mobile's MAC address on BLE devices using TI chips. TI is a leading BLE chip manufacturer, which implements multiple security levels for its widely used BLE chips. 
We find two critical vulnerabilities in TI's BLE SDK.
%, which is broadly used CC26XX \cite{CC2640-Top}:

First, the TI's BLE SDK incorrectly implements the Secure Connections Only mode.
TI's SDK allows an application to set a Secure Connections Only mode flag as true or false. However, the related code only checks if the incoming pairing request enables the {\em Secure Connections (SC)} bit and does not check if the negotiated pairing protocol is the Passkey Entry or Numerical Comparison protocol. The SC bit only refers to if a mobile or its peer BLE device supports BLE 4.2 and 5.x.  However, their BLE Stack SDK provides the GAP Bong Manager layer (re: “related code”) in source code format which allows the application to inspect and perform accordingly based on the peer IO Capabilities field.

Second, the TI's BLE stack caches and reuses the LTK's property for a bonded mobile. A LTK can be an unauthenticated-and-no-MITM-protection key created by Just Works or an authenticated-and-MITM-protection key created by Passkey Entry, Numeric Comparison or OOB. 
Assume that a victim mobile uses secure pairing to pair with a victim BLE device based on TI chips and generate an authenticated-and-MITM-protection LTK. If a fake mobile with the victim mobile's MAC address uses Just Works and pairs with the victim device, the generated LTK still has the property of authenticated-and-MITM-protection. Therefore, the fake mobile can access attributes with the authenticated read/write permission. We have tested and proved the vulnerabilities on TI's CC2640, CC2640R2F, and CC2650, and reported the identified vulnerabilities to TI and a patched SDK was released recently~\cite{Ti-new-SDK,Tifix}.

\end{document}